\documentclass[12pt,reqno]{amsart}
\usepackage{amsthm,amsfonts,amssymb,euscript}

\def\emph#1{{\it #1}}
\def\textbf#1{{\bf #1}}
\def\Gab{\underline{\Ga}}
\newcommand{\bea}{\begin{eqnarray}}
\newcommand{\eea}{\end{eqnarray}}
\def\beaa{\begin{eqnarray*}}
\def\eeaa{\end{eqnarray*}}

\def\Lieh{\widehat{\Lie}}

\def\piL  {\, ^{(L)}\pi}
\def\piLb {\, ^{(\Lb)}\pi}

\def\ub{{\underline{u}}}

\def\Dcal{{\mathcal D}}

\def\II{{\mathcal I}}
\def\ba{\begin{array}}
\def\ea{\end{array}}
\def\be#1{\begin{equation} \label{#1}}
\def \eeq{\end{equation}}

\newcommand{\nn}{\nonumber}

\def\pih{\hat{\pi}}
\def\nn{\nonumber}

\def\Lie{{\mathcal L}}
\def\tr{\mbox{tr}}

\def\RR{{\mathcal R}}

\def\Pib{\underline{\Pi}}

\def\a{\alpha}
\def\alphab{{\underline\alpha}}
\def\b{\beta}

\def\thb{\underline{\th}}
\def\Gab{\underline{\Ga}}
\def\aa{\alphab}
\def\bb{\betab}

\def\ga{\gamma}
\def\Ga{\Gamma}
\def\de{\delta}
\def\De{\Delta}
\def\ep{\epsilon}

\def\La{\Lambda}
\def\si{\sigma}

\def\om{\omega}
\def\omegab{{\underline\omega}}
\def\Om{\Omega}
\def\ro{\rho}

\def\th{\theta}

\def\ze{\zeta}
\def\nab{\nabla}

\def\bb{\underline{\b}}

\newcommand{\trchb}{\tr \chib}
\newcommand{\wtrchb}{\widetilde{\tr \chib}}
\newcommand{\chih}{\hat{\chi}}
\newcommand{\chib}{\underline{\chi}}

\newcommand{\etab}{\underline{\eta}}
\newcommand{\chibh}{\underline{\hat{\chi}}\,}
\newcommand{\omb}{{\underline{\om}}}
\def\f14{\frac{1}{4}}
\def\f12{{\frac{1}{2}}}

\def\c{\cdot}

\newcommand{\les}{\lesssim}

\def \piO{\,^{  (O)  }     \pi\,}

\def\Lieh{\widehat{\Lie}}

\newcommand{\Tr}{\text{Tr}\,}

\newcommand{\QQ}{\mathcal{Q}}

\newcommand{\DD}{{\mathcal D}}

\def\RRb{\underline{\mathcal R}}

\def\OO{{\mathcal O}}
\def\OS{\,\,^{(S)}\OO}
\def\OH{\,^{(H)}\OO}

\def\RRLa{\,^{(\La)}\RR}
\def\RRbLa{\,^{(\La)}\RRb}
\def\RRLasup{\,^{[\La]}\RR}
\def\RRbLasup{\,^{[\La]}\RRb}

\def\fLa{\,^{(\,\La)}f}
\def\fLat{\,^{(\,\widetilde{\La})}f}
\def\Lat{{\widetilde \La}}

\def\RRold{\dot{\RR}}
\def\RRbold{\dot{\RRb}}

\def\OOold{\dot{\OO}}
\def\scold{\dot{\sc}}

\def\PiO{\,^{(O)}\Pi  }

\def\Lat{\tilde{\La}}

\def\Lsc{{\mathcal L}_{(sc)}   }
\def\Lscold{{\mathcal L}_{(\scold)}   }

\def\ub{\underline{u}}

\def\Lb{{\underline{L}}}

\def\sc{{\text sc}}
\def\QQ{{\mathcal Q}}

\def\Hb{{\underline{ H}}}
\def\pr{\partial}

\def\chih{\hat{\chi}}
\def\trch{\mbox{tr}\chi}

\def\trchbt{\widetilde{\trchb}}

\def\sgn{{\text sgn}}

 \def\Trsc{Tr_{(sc)}}

\usepackage{graphics}
\usepackage{graphicx}
\begin{document}
\theoremstyle{plain}
  \newtheorem{theorem}[subsection]{Theorem}
  \newtheorem{conjecture}[subsection]{Conjecture}
  \newtheorem{proposition}[subsection]{Proposition}
  \newtheorem{lemma}[subsection]{Lemma}
  \newtheorem{corollary}[subsection]{Corollary}

\theoremstyle{remark}
  \newtheorem{remark}[subsection]{Remark}
  \newtheorem{remarks}[subsection]{Remarks}

\theoremstyle{definition}
  \newtheorem{definition}[subsection]{Definition}

\include{psfig}

\author{Sergiu Klainerman}
\address{Department of Mathematics, Princeton University,
 Princeton NJ 08544}
\email{ seri$@$math.princeton.edu}
\title[Trapped surfaces]{On emerging scarred  surfaces  for the Einstein vacuum equations}

\author{Igor Rodnianski}
\address{Department of Mathematics, Princeton University, 
Princeton NJ 08544}
\email{ irod$@$math.princeton.edu}
\subjclass{35J10\newline\newline
}
\vspace{-0.3in}
\maketitle
\maketitle
\section{introduction} This is a follow up on our work \cite{K-R:trapped}      
in which  we have presented a modified, simpler  version of the remarkable recent result of Christodoulou, see  \cite{Chr:book},  on  the  formation     of  evolutionary trapped   surfaces in vacuum.  The approach   in  \cite{K-R:trapped},  based on a different scaling\footnote{The natural   null, parabolic,  scaling of the Einstein vacuum equations}  than that of    \cite{Chr:book},  allowed us   not only  to reprove Christodoulou's 
  trapped surface  result,     but also enabled us  to   localize
 with respect to   small  angular regions.
    This  led us, in particular,    to  a  simple result concerning the  formation   of pre-scarred 
  surfaces\footnote{These are surfaces for which the outgoing expansion
  is negative in an open subset of the  surface}. Both  results were
  based on the proof of a   semi-global existence  theorem which
  established the propagation of  precise  estimates, for both  curvature
  and Ricci coefficients,  starting with non-trivial initial conditions on an outgoing null
  hypersurface.

In this paper we provide a   considerable extension of  our result on pre-scared surfaces to allow  for  the formation of a surface with multiple pre-scared angular regions which, together, can cover an arbitrarily large portion of the surface. In a forthcoming paper we plan to show    that  once a significant part of the surface is pre-scared, it can be additionally 
deformed to produce a bona-fide trapped surface. This result implies, in particular, that Christodoulou's  crucial 
 uniform lower bound  initial  condition necessary 
 for the formation of a trapped surface can be relaxed 
 to an average condition, which requires  only that the lower bound
 holds true  only on  a sufficiently large angular  portion of the 
 initial outgoing  null hyper-surface. 

In this paper we state and discuss three related results.
\begin{enumerate}
\item We  state an optimal propagation result, critical with respect to the natural  null scaling  of   Einstein vacuum equations introduced in   \cite{K-R:trapped} (which dealt   with the subcritical regime), see
 theorem \ref{thm:main}.
  In this paper, prompted, in part,  by   our interest  on  pre-scarred surfaces    and in part by
reflecting on the scale transformation  in the work of Reiterer and
Trubowitz \cite{R-T},  we   note      that  the  argument of the main propagation
theorem in \cite{K-R:trapped}   proves in fact  a stronger,   indeed   optimal 
 result.  We are happy to acknowledge that a related  result is  stated 
  in  theorem 8.1. of  \cite{R-T}, in a different setting. We would like to thank Reiterer and Trubowitz  for drawing our attention  and  making an effort  to explain  its formulation  to us.   
 \item We state, see theorem \ref{thm.main.loc},   
  an angular localized version of   the global  energy estimates
 for the null  curvature  components of theorem \ref{thm:main}. 
 The proof relies on a natural  modification of the 
   proof in theorem  \ref{thm:main} and is  discussed in  section \ref{sect:loc.curv}.
 
 \item  We give a large class of critical, sufficient  conditions on the initial data,
 which lead to the  formation of pre-scarred surfaces.
 The main result is stated in theorem \ref{thm:mainII}. 
 The proof rests on theorem \ref{thm:main} as well
 as on a localized version of the  Ricci coefficient 
 estimates in \cite{K-R:trapped}.  As mentioned above,
  the importance  of this result is due to the fact that 
 once a significant part of a surface is pre-scarred,
 it can be deformed to  a real  trapped surface.

\end{enumerate}
 
Concerning the new propagation result  stated in  theorem \ref{thm:main},
we  note that  the main new idea is to use, in addition
 to the small  parameter $\de>0$, originating in the short pulse method
 of   \cite{Chr:book},   a  new  small parameter $\ep$ with  $\de^{1/2}\ep^{-1}$
 sufficiently small.  The parameter
 $\de$  is used to define scale invariant norms, similar to those
 we have introduced in \cite{K-R:trapped}  but with one
  important modification. In the main result of  \cite{K-R:trapped}, for example, the scaling 
   was such that   all  null curvature components, except the component 
    denoted by  $\a$,  
   were bounded  (in  its scale invariant norms) . The behavior (in  the scale invariant norm) of the \textit{anomalous} component $\a$, on the other
   hand, was $\de^{-1/2}$. Here we  choose  the scaling with respect
   to $\de $ such that    the scale invariant norm of $\a$ is bounded,
   independent of the second parameter $\ep$,   and  the scale invariant norms of  all other curvature components  are  proportional  to   $\ep$,  i.e. small.  All results in \cite{K-R:trapped} correspond
   precisely  to the case when $\ep$ is chosen to be proportional to $\de^{1/2}$.
   It is quite  remarkable that the proof of the stronger propagation result 
   in theorem \ref{thm:main}    is    exactly  the same as in \cite{K-R:trapped}.  
   This is surprising, especially considering that the initial data in 
   theorem \ref{thm:main} is allowed to be $\de^{-\frac 12} \ep$ times bigger\footnote{More precisely all components of the curvature tensor , except $\a$,
   are  $\de^{-\frac 12} \ep$ times bigger. The $\a$ component 
   behaves exactly the same   as in    \cite{K-R:trapped}. } than 
   that in \cite{K-R:trapped} (as measured in absolute, unscaled norms). In \cite{K-R:trapped} 
   nonlinear non-anomalous interactions were controlled by the scale invariant H\"older estimates
   $$
   \|\psi\c\phi\|_{\Lscold^p} \les \de^{\frac 12} \|\psi\|_{\Lscold^r} \|\phi\|_{\Lscold^q},\qquad \frac 1r+\frac 1q=\frac 1p.
   $$
   In this work the new critical scaling does not generate a small factor of $\de^{\frac 12}$ in such interactions. Instead
   we have 
   $$
   \|\psi\c\phi\|_{\Lsc^p} \les  \|\psi\|_{\Lsc^r} \|\phi\|_{\Lsc^q},\qquad \frac 1r+\frac 1q=\frac 1p.
   $$  
 For non-anomalous $\psi$ and $\phi$ the scale invariant norms on the right hand side are both 
 of size $\ep$ and so is the expected value of the left hand side norm. This analysis indicates that 
 with the new scaling the factor $\de^{\frac 12}$ of quadratic interactions is effectively replaced by the independent small    
   parameter $\ep$.

In the result on the formation of a pre-scarred surface we describe a set of initial data which lead to a
space-time with a surface containing approximately $\de^{-\frac 12} q$ angular regions  of size $\de^{\frac 12}q^{-1}$,  each of which is
pre-trapped   for some sufficiently small parameter $q$.
\vskip 1pc
   
    We start by recalling the framework
 of   double
  null foliations in which   the results of both  \cite{Chr:book} and \cite{K-R:trapped}  are  formulated.

\subsection{Double null foliations} \label{subsec:doublenull}
 We consider a region $\DD=\DD(u_*,\ub_*)$ of a vacuum spacetime $(M,g)$
    spanned by a double null foliation  generated by the optical functions  $(u,\ub)$  increasing towards the future,   $0\le u\le u_*$ and $0\le\ub\le  \ub_*$.  We denote by $H_u$ the outgoing  null hypersurfaces generated by the  level surfaces of $u$  and     by  $\underline{H}_{\ub}$ the incoming  null hypersurfaces generated  level hypersurfaces  of  $\ub$.  We write 
$S_{u,\ub}=H_u\cap \underline{H}_{\ub}$ and   denote by $H_{u}^{(\ub_1,\ub_2)}$,  and $\underline{H}_{\ub}^{(u_1,u_2)}$ the regions of these null hypersurfaces   defined by $\ub_1\le\ub\le\ub_2$ and respectively $u_1\le u\le u_2$.
    Let $L,\Lb $ be the geodesic vectorfields associated to the two foliations and  define the null lapse $\Om$ and connection, or Ricci, coefficients,
    $\chi, \om, \eta, \etab, \chib, \omb$,
\bea
\frac 1 2 \Omega^2=-g(L,\Lb)^{-1}\label{eq:def.omega}
\eea
\begin{equation}
\begin{split}
&\chi_{ab}=g(D_a e_4,e_b),\, \,\,\, \quad \chib_{ab}=g(D_a e_3,e_b),\\
&\eta_a=-\frac 12 g(D_3 e_a,e_4),\quad \etab_a=-\frac 12 g(D_4 e_a,e_3)\\
&\omega=-\frac 14 g(D_4 e_3,e_4),\quad\,\,\, \omegab=-\frac 14 g(D_3 e_4,e_3)
\end{split}
\end{equation}
where  $
e_3=\Omega\Lb,\, e_4=\Omega L$
 and    $D_a=D_{e_{(a)}}$. As usual we decompose the null second fundamental forms $\chi, \chib$ into their traceless parts $\chih, \chibh$
 and traceless parts, or \textit{expansions}, $\trch, \trchb$. 
 We also introduce the  null curvature components,
 \begin{equation}
\begin{split}
\a_{ab}&=R(e_a, e_4, e_b, e_4),\quad \, \,\,   \aa_{ab}=R(e_a, e_3, e_b, e_3),\\
\b_a&= \frac 1 2 R(e_a,  e_4, e_3, e_4) ,\quad \bb_a =\frac 1 2 R(e_a,  e_3,  e_3, e_4),\\
\rho&=\frac 1 4 R(Le_4,e_3, e_4,  e_3),\quad \si=\frac 1 4  \,^*R(e_4,e_3, e_4,  e_3)
\end{split}
\end{equation}
Here $\, ^*R$ denotes the Hodge dual of $R$.  We denote by $\nab$ the 
induced covariant derivative operator on $S(u,\ub)$ and by $\nab_3$, $\nab_4$
the projections to $S(u,\ub)$ of the covariant derivatives $D_3$, $D_4$.
We note the formulas,
\begin{equation}
\begin{split}
&\omega=-\frac 12 \nab_4 (\log\Omega),\qquad \omegab=-\frac 12 \nab_3 (\log\Omega),\qquad
\eta +\etab =2\nab  (\log\Omega)
\end{split}
\end{equation}
We recall also the formula  for the Gauss curvature $K$ of $S(u,\ub)$,
\bea
K&=&-\rho+\frac 1 2 \chih\c\chibh -\frac 1 4 \trch \c\trchb
\eea

As well known,  our  space-time slab  $\DD(u_*, \ub_*)$  is completely 
determined  (for small values of $u_*, \ub_*$)   by  specifying, \textit{freely},   the traceless parts of the null second fundamental forms $\chih$, respectively $\chibh$, along     the null, characteristic,  hypersurfaces $H_0$,  respectively  $\Hb_0$,  corresponding to
 $\ub=0$, respectively $u=0$, and prescribing $\trch$ together with $\trchb$ on $S(0,0)$.
Following \cite{Chr:book} we assume that   our  data is trivial along 
$\Hb_0$, i.e. assume that $H_0$ extends for $\ub<0$   and  the  spacetime    $(M, g)$ is Minkowskian for   $\ub <0$  and all values of  $u\ge 0$. Moreover we can construct our double null foliation such that
 $\Om=1$ along $H_0$, i.e.,
\bea
\Om(0,\ub)=1, \qquad 0\le \ub\le\ub_*.
\eea
We  also introduce the notation,
\bea
   \trchbt=\trchb-\trchb_0, \qquad  \trchb_0=-\frac{4}{\ub-u+2r_0}
   \eea
 where $\trchb_0$ is the  flat value of $\trchb$ along the  initial hypersurface $\Hb_0$. We denote by $\ga$ the induced metric on the surfaces  $S(u,\ub)$
 of intersection between $H_u$ and $H_{\ub}$.  A space-time tensor
 tangent to $S(u, \ub)$ is called an $S-$ tensor, or horizontal tensor.
 
 We define  systems of, local, transported coordinates along  the null 
 hypersurfaces $H$ and $\Hb$.  Starting with a local coordinate system $\th=(\th^1, \th^2)$
on   $U\subset  S(u,0)\subset H_u$, we parametrize any point along the null geodesics starting
 in $ U$ by the  the corresponding coordinate $\th$ and affine parameter $\ub$. Similarly,
 starting with a local coordinate system $\thb=(\thb ^1, \thb^2)$  on 
 $V\subset S(0,\ub)\subset \Hb_{\ub}$
 we parametrize  any point   along the null geodesics starting
 in $ V$ by the  the corresponding coordinate $\thb$ and affine parameter $u$.

\subsection{Signature}  To every  null curvature
component $\a,\b,\rho,\si, \bb,\aa$,   null Ricci coefficients
components $\chi,\ze,\eta, \etab, \om,\omb$, and metric $\ga$  we assign a signature  according to the following rule:
\bea
\sgn(\phi)=1\c  N_4(\phi)+\frac 12 \c N_a(\phi)+0\c N_3(\phi)    - 1
\eea
where $N_4(\phi), N_3(\phi), N_a(\phi)$ denote  the number of  times $e_4$,
respectively $e_3$ and $(e_a)_{a=1,2}$, which  appears in the
definition of $\phi$.
 Thus,
\beaa
\sgn(\a)=2,\quad  \sgn (\b)=1+1/2, \quad  \sgn(\rho, \si)=1,\quad  \sgn(\bb)=1/2,\quad 
\sgn(\aa)=0.
\eeaa
Also,
\beaa
\sgn(\chi)=\sgn( \om)=1,\quad \sgn(\ze, \eta, \etab)=1/2,\quad \sgn(\chib)=\sgn(\omb)=\sgn(\ga)= 0.
\eeaa
Consistent with this definition we have, for any given null component $\phi$,
\beaa
\sgn(\nab_4\phi)=1+\sgn(\phi),\quad \sgn(\nab\phi)=\frac 1 2 +\sgn(\phi),\quad
\sgn(\nab_{3}\phi)=\sgn(\phi).
\eeaa
Also, based on our convention,
\bea
\sgn(\phi_1\cdot\phi_2)=\sgn (\phi_1)+\sgn(\phi_2).
\eea
\subsection{Main  equations}
 As in \cite{K-R:trapped} we  denote all  Ricci coefficients  $\{ \chi, \om, \eta, \etab, \trchbt, \chibh, \omb\}$  by  $\psi^{(s)}$, with $s$ the signature of the specific component. We further differentiate between the components
 $\psi^{(s)}_4\in\{\chi, \eta, \omb\} $, which verify transport equations in the $e_4$ direction,
 and $\psi_3^{(s)}\in \{\om, \etab, \trchbt,\chibh\} $ which verify transport equations in the $e_3$ direction.   We  denote by $\Psi^{(s)}$ the 
 null curvature  components of signature $s$.  With these notation
 the null structure equations, see precise equations in   section 3 of \cite{K-R:trapped},   take the form,
 \bea
\nab_4 \psi_4^{(s)}&=&\sum_{s_1+s_2=s+1}\psi^{(s_1)}\c\psi^{(s_2)}+\Psi^{(s+1)}\label{eq:transp.symb4}\\
\nab_3 \psi_3^{(s)}&=& \trchb_0\c\psi_3^{(s)}+    \sum_{s_1+s_2=s}\psi^{(s_1)}\c\psi^{(s_2)}+\Psi^{(s)}
\label{eq:transp.symb3}
\eea
Similarly we write the null Bianchi identities in the from,
\bea
\nab_4\Psi^{(s)}_4&=&\nab \Psi^{(s+\frac 1 2 )}+\sum_{s_1+s_2=s+1} \psi^{(s_1)}\c\Psi^{(s_2)}\\
\nab_3\Psi^{(s)}_3&=&\nab \Psi^{(s-\frac 1 2 )}+\sum_{s_1+s_2=s} \psi^{(s_1)}\c\Psi^{(s_2)}
\eea
where $\Psi_4\in\{\a,\b,\rho, \si\}$ and $\Psi_3\in\{\b,\rho,\si, \bb,\aa\}$.

\subsection{Scale invariant norms}
 
 For any horizontal tensor-field $\psi$  with signature  $\sgn(\psi)$ we define the following 
 scale invariant norms along the null hypersurfaces $H=H_u^{(0,\de)}$ and $\underline{H}=\underline{H}_{\ub}^{(0,1)}$.
 \bea
\|\psi\|_{\Lsc^2(H)}&=&\de^{\sgn(\psi)-1} \|\psi\|_{L^2(H)},\quad 
\|\psi\|_{\Lsc^2(\underline{H})  }=\de^{\sgn(\psi)-\frac 1 2 } \|\psi\|_{L^2(\underline{H})}
\eea
We also define  the scale invariant norms on the $2$ surfaces $S=S_{u,\ub}$,
\bea
\|\psi\|_{\Lsc^p(S)}&=&\de^{\sgn(\psi)-\frac 1 p } \|\psi\|_{L^p(S)}
\eea
We have,
\bea
\|\psi\|_{\Lsc^2(H_u^{(0,\ub)})}^2   &=&\de^{-1} \int_0^{\ub}\|\psi\|_{\Lsc^2(u,\ub')} ^2d\ub',\qquad 
\|\psi\|_{\Lsc^2(\Hb_{\,\ub}^{(0,u)})}^2=\int_0^{u}\|\psi\|_{\Lsc^2(u',\ub)}^2 du'
\eea
We  denote  the scale invariant  $L^\infty$  norm in $\DD$   by  $ \|\psi\|_{\Lsc^\infty}$.  
\begin{remark}
\label{rem:oldnewscales}
These norms correspond to a different scaling than that introduced in  \cite{K-R:trapped}. Indeed  in \cite{K-R:trapped}  the scale invariant norms were based on  the definition of the scale of  an horizontal component
 of scale   $\sc(\psi)=-\sgn(\psi)+\frac 1 2 $. The norms introduced here 
 would correspond to a  new definition of scale give by  
  $\sc(\psi)=-\sgn(\psi)$.  To distinguish between them we denote
  the old scaling by $\scold$.  Thus, for example,
  \beaa
  \|\psi\|_{\Lsc^p(S)}=\de^{-1/2}  \|\psi\|_{\Lscold^p(S)}
  \eeaa
\end{remark}
\begin{remark}
With the new scale invariant norms introduced here we have,
\bea
\|\psi_1\c\psi_2\|_{\Lsc^2(S)} &\les& \|\psi_1\|_{\Lsc^\infty(S)}\c \|\psi_2\|_{\Lsc^2(S)}
\eea
or,
\bea
\|\psi_1\c\psi_2\|_{\Lsc^2(H)} &\les& \|\psi_1\|_{\Lsc^\infty(H)}\c \|\psi_2\|_{\Lsc^2(H)}\label{product.inv.estim}
\eea
These differ from the situation in \cite{K-R:trapped} where the corresponding estimates (with $(\sc)$ replaced by $(\scold)$)  had an additional power of $\de^{1/2}$ on the right. 
\end{remark}
\textit{Curvature norms.}\, We  introduce our main curvature norms
 \begin{equation}
 \label{RR-norms}
 \begin{split}
  \RR_0(u,\ub):&=\|\a\|_{\Lsc^2(H_u^{(0,\ub)})}+\RR_0'(u,\ub')\\
 \RR_0'(u,\ub'):&=   \ep^{-1} \|(\b,\rho,\si,\bb,K)\|_{\Lsc^2(H_u^{(0,\ub)})}\\
   \RR_1(u,\ub):&= \|\nab_4\a\|_{\Lsc^2(H_u^{(0,\ub)})}+\RR_1'(u,\ub)\\
\RR_1'(u,\ub):&=    \ep^{-1}\|\nab( \a, \b,\rho,\si, \bb,K)\|_{\Lsc^2(H_u^{(0,\ub))})}\\
   \\
    \RRb_0(u,\ub):&=\|\b\|_{\Lsc^2(\Hb_{\ub}^{(u,0)})}+\RRb'_0(u,\ub')\\
  \RRb'_0(u,\ub)&=  \ep^{-1}\|(\rho,\si,\bb,\aa,K)\|_{\Lsc^2(\Hb_{\ub}^{(0,u)})}\\
   \RRb_1(u,\ub):&= \|\nab_3\aa\|_{\Lsc^2(\Hb_{\ub}^{(u,0)})}+ \RRb'_1(u,\ub)\\
   \RRb'_1(u,\ub): &=   \ep^{-1}      \|\nab(  \b,\rho,\si, \bb,\aa,K)\|_{\Lsc^2(\Hb_{\ub}^{(0,u)})}
   \end{split}
    \end{equation}
      Also,
     \begin{equation}
 \begin{split}
 \RR=\RR_0+\RR_1,\qquad \RRb=\RRb_0+\RRb_1\label{RR-norms.f}
  \end{split}
    \end{equation}
    \begin{remark}
We have included the Gauss curvature $K$ with the null components. Since  $
K=-\rho+\frac{1}{2}\chih\c\chibh -\frac 1 4 \trch\trchb
$  we  easily deduce that, 
$$
\ep^{-1} \|K\|_{\Lsc^2(H_u^{(0,\ub)})}\les \ep^{-1} \|\rho\|_{\Lsc^2(H_u^{(0,\ub)})}+\big(1+(\ep^{-2}\de)^{\frac 12}\big)\OS_{0,\infty}\OS_{0,2}
.$$
\end{remark}

      \begin{remark}
    All  curvature norms above   have a factor of $\ep^{-1}$
    in front of them except for 
       $\|\a\|_{\Lsc^2(H_u^{(0,\ub)})}$, $\|\nab_4\a\|_{\Lsc^2(H_u^{(0,\ub)})}$  and     $\|\b\|_{\Lsc^2(\Hb_{\ub}^{(u,0)})}$.    These correspond  exactly 
        to the      \textit{anomalous} curvature norms  of \cite{K-R:trapped}.
         \end{remark}  
    To rectify the anomaly of $\a$ we introduce, as in  \cite{K-R:trapped},   an additional scale-invariant norm, 
$$
\RR_0^{(\ep)}[\a](u,\ub):= \sup_{^{(\ep)} H\subset H} \ep^{-1}\|\a\|_{\Lsc^2(^{(\ep)} H)},
$$
where $^{(\ep)} H$ is a piece of the hypersurface $H=H_u^{(0,\de)}$ obtained by evolving an angular  disc $S_\ep\subset S_{u,0}$ 
of radius $\ep$ relative  to our transported coordinates.
We define the initial  quantity  $\RR^{(0)}$ by,
\bea
\RR^{(0)}=\sup_{0\le \ub\le \de}\big( \RR(0, \ub)+\RR_0^{(\ep)}[\a](0,\ub)\big)
\label{RR-initial}
\eea
\subsection{Connection coefficients norms}\,   
    We introduce the  Ricci coefficient norms, with the supremum taken over all
    surfaces $S=S(u',\ub'), 0\le u'\le u,\, 0\le \ub'\le \ub$,
    \begin{equation}
    \begin{split}
 \OS_{0,\infty}(u,\ub)&=\ep^{-1} \sup_{S}\|(\chih,\om ,\eta,\etab,\trchbt,\chibh, \omb)\|_{\Lsc^\infty(S)}\\
  \OS_{0,2}(u,\ub)&= \sup_{S}\big(\|\chih\|_{\Lsc^2(S) } +\|\chibh\|_{\Lsc^2(S)}\big)+ \OS'_{0,2}(u,\ub)\\
   \OS'_{0,2}(u,\ub)&=
\ep^{-1} \sup_{S}\|(\trch, \om,\eta,\etab,\trchbt, \omb)\|_{\Lsc^2(S)}\\
 \OS_{0,4}(u,\ub)&=\ep^{-1/2} \sup_{S}\big( \|\chih\|_{\Lsc^4(S) }
 +\|\chibh\|_{\Lsc^4(S)}\big)+ \OS'_{0,4}(u,\ub)\\
 \OS'_{0,4}(u,\ub) &=\ep^{-1} \sup_{S}\|(\trch, \om,\eta,\etab,\trchbt, \omb)\|_{\Lsc^4(S)}
 \\
  \OS_{1,4}(u,\ub)&=\ep^{-1} \sup_{S}\|\nab(\chi, \om,\eta,\etab,\trchbt, \chibh, \omb)\|_{\Lsc^4(S)}\\
   \OS_{1,2}(u,\ub)&=\ep^{-1} \sup_{S}\|\nab(\chi, \om,\eta,\etab,\trchbt, \chibh, \omb)\|_{\Lsc^2(S)}\\
  \OH(u,\ub)&=\ep^{-1} \|\nab^2(\chi, \om,\eta,\etab,\trchbt, \chibh, \omb)\|_{\Lsc^2(H_u^{(0,\ub)})}  
  \end{split}
    \end{equation}
    and,
   \bea
   \OO&=&\OS_{0,2}+\OS_{0,4}+\OS_{0,\infty}+\OS_{1.4}+\OH \label{OO-norms.f}
  \eea
  \begin{remark}
  Note that the only  norms  which do not contain powers of $\ep^{-1}$
  are the $\Lsc^2(S)$ norms  of $\chih$ and $\chibh$. This anomaly
  is also manifest in the $\Lsc^4(S)$ norms of the same quantities. These
 are precisely the same    quantities which were anomalous in \cite{K-R:trapped}, with respect  to the  $\scold$ scaling.
    \end{remark}
To cure the above anomaly  we define the auxiliary  norms,
 $$
  \OS^{(\ep)}_{0,4}(u,\ub)=\ep^{-1} \sup_S   \sup_{S_\ep\subset S} 
  \|(\chih,\chibh)\|_{\Lsc^4(S_\ep)}
 $$
 with $S_\ep$ - an angular subset of $S$ of size $\ep$ relative to our transported coordinates.
 
 Finally we define the initial  data quantity:
 \bea
 \OO^{(0)}=\sup_{0\le \ub \le \de} \big(\OO(0,\ub)+ \OS^{(\ep)}_{0,4}(0,\ub)\big)
 \eea

 \subsection{Initial conditions} 
  Define the main  initial data quantity,
 \begin{equation}
 \begin{split}
& \II^{(0)}(\ub)= 
  \sum_{0\le k\le 2}  \|\nab_4^k \chih_0\|_{\Lsc^2(0,\ub)}\\
 &+ \ep^{-1}\bigg( \|\chih_0\|_{\Lsc^\infty(0,\ub)}+
  \sum_{0\le k\le 1} \, \sum_{1\le m\le 4}   \|\nab^{m-1}  \nab_4 ^k\, \nab \chih_0\|_{\Lsc^2(0,\ub)}\bigg)
    \end{split}
          \end{equation}
        
            or, in the natural norms,          
 \beaa
&& \II^{(0)}(\ub)= 
  \sum_{0\le k\le 2} \de^{k+1/2} \|\nab_4^k \chih_0\|_{L^2(0,\ub)}   \\
      &&+
      \ep^{-1}\bigg(\de\|\chih_0\|_{L^\infty(0,\ub)}+      \sum_{0\le k\le 1} \, \sum_{1\le m\le 4} \de^{\frac{m+1}{2}+k}  \|\nab^{m-1}  \nab_4 ^k\, \nab \chih_0\|_{L^2(0,\ub)}\bigg)   
      \eeaa    
 \subsection{Main propagation  result} 
The first result establishes 
 the boundedness of the initial curvature and Ricci coefficent
 scale invariant norms $\RR^{(0)}$,  $\OO^{(0)}$   in terms of $\II^{(0)}$.
  \begin{proposition}
           \label{thm.main.initial}
 Assume that the initial data along
          $\Hb_0$ is flat and that   $\II^{(0)}<\infty$
          along $H_0^{(0,\de)}$. Then, for $\de^{1/2} \ep^{-1}$ and $\ep>0$  sufficiently small
   we have,  with $C$ a     fixed  super-linear polynomial 
    \bea
    \RR^{(0)}+ \OO^{(0)} \les \II^{(0)} +C(\II^{(0)})
    \label{eq:initialOR}
    \eea
   Also, starting with $\RR^{(0)}<\infty$ and $\de^{1/2} \ep^{-1}$, $\ep$ sufficiently small,       we have,  with  $C$ a  fixed  super-linear polynomial,
    \bea
    \OO^{(0)}&\les& \RR^{(0)}+C(\RR^{(0)})
    \eea
    \end{proposition}     
 We can now state our main  propagation result.
 \begin{theorem}[Main Theorem I]  \label{thm:main}  Under the assumption $\RR^{(0)}<\infty$,   if    $\de^{1/2}\ep^{-1}$ and $\ep$ are   sufficiently small then,
 for  $0\le u\le 1$, $0\le\ub\le \de$,  with  $C$ a  fixed  super-linear polynomial,
 \beaa
 (\RR+\RRb+\OO )(u,\ub)\les \RR^{(0)}+C(\RR^0)
 \eeaa

 \end{theorem}
    {\bf Remark 1}.\,  The results presented extends  all   the results 
 of \cite{K-R:trapped}. Indeed, to derive the results of propositions 2.5,  theorems 2.6, and  2.7   there,  it suffices to  choose $\ep=\mu \de^{1/2} $ with $\mu$  sufficiently small.  
 
 {\bf Remark 2}.\,
  The additional  smallness assumption  on   $\de^{1/2}\ep^{-1}$ 
  is due to the  lower order terms     which appear in  some of the calculus inequalities  presented in the next section. 
  
 In the remaining part of this section we introduce norms for the deformation tensors of the geodesic null  generators   $L, \Lb$ and rotation vectorfields
 $O$ and give a short sketch of the proof of theorem \ref{thm:main}.
  
 \subsection{Deformation tensors norms for $L,\Lb$}

   If $\pi$ is the deformation tensor
    of either $L$ or $\Lb$  we denote by $\pi^{(s)}$ its  null component of
    of signature $s$.  We now introduce the norms for $\piL$ and $\piLb$
    as follows,
    \bea
    \Pi_0=\Pi_{0,4}+\Pi_{0,\infty},\qquad \Pib_0=\Pib_{0,4}+\Pib_{0,\infty}
    \eea
    with,
    \begin{equation}
    \begin{split}
    \Pi_{0,4}&= 
\ep^{-1} \sum_{s\in\{0,\frac 12\}} \|\piL^{(s)}\|_{\Lsc^4(S)}+\ep^{-\frac 12} \|\piL^{(1)}\|_{\Lsc^4(S)},\\
\Pi_{0,\infty}&=\ep^{-1} \sum_{s\in\{0,\frac 12, 1\}} \|\piL^{(s)}\|_{\Lsc^\infty(S)},\\
\Pib_{0,4}&=
\ep^{-1} \sum_{s\in\{\frac 12, 1\}} \|\piLb^{(s)}\|_{\Lsc^4(S)}+\ep^{-\frac 12} \|\piLb^{(0)}\|_{\Lsc^4(S)},\\
\Pib_{0,\infty}&=\ep^{-1} \sum_{s\in\{0,\frac 12, 1\}} \|\piLb^{(s)}\|_{\Lsc^\infty(S)} 
\end{split}
\end{equation}
  We introduce also the first derivative norms,
    \begin{equation}
    \begin{split}
  \Pi_1&=\|\nab_4\piL^{(0)}\|_{\Lsc^4(S)}+\sum_{s\in\{\frac 12,1\}} 
\ep^{-1}\|\bar\nab\piL^{(s)}\|_{\Lsc^4(S)},\\
\Pib_1&=\|\nab_4\piLb^{(0)}\|_{\Lsc^4(S)}+\|\nab_3\piLb^{(0)}\|_{\Lsc^4(S)}\\
&+
\ep^{-1}\|\bar\nab\piLb^{(\frac 12)}\|_{\Lsc^4(S)} + \ep^{-1}\|(\nab,\nab_3)\piLb^{(1)}\|_{\Lsc^4(S)} ,  
  \end{split}
\end{equation}  
We also set,
\beaa
\Pi=\Pi_0+\Pi_1,\qquad \Pib=\Pib_0+\Pib_1
\eeaa
    \subsection{Deformation tensor norms for $O$}
  We recall   the  rotation vectorfields $^{(i)} O$ obeying the commutation relations 
$$
[^{(i)}O,^{(j)}O]=\in_{ijk}\, ^{(k)}O,
$$
 were  defined, see section 13 in  \cite{K-R:trapped},  by parallel transport starting with  the standard rotation vectorfields on ${\Bbb S}^2=S_{u,0}\subset H_{u,0}$ 
along the integral curves of $e_4$. Suppressing the index $^{(i)}$ we have, 
\bea
\nab_4 O_b=\chi_{bc} O_c.
\eea
The only non-trivial components of the deformation tensor $\pi_{\a\b}=\frac 12 (\nab_\a O_\b+\nab_\b O_\a)$ 
are given below:
\begin{align*}
&\pi_{34}=-2(\eta+\etab)_a O_a,\\
&\pi_{ab}=\frac 12 (\nab_a O_b+\nab_b O_a)=\frac 12(H_{ab}+H_{ba}),\\
&\pi_{3a}=\frac 12 (\nab_3 O_a-\chib_{ab} O_b):=\frac 12 Z_a.
\end{align*}
The quantities, $H$ and $Z$ can be assigned signature and scaling, (consistent with those for the Ricci coefficients and curvature components) according to.  
 \bea
 sgn(H)=0,\qquad sgn(Z)=-\frac 12.
 \eea
 Similarly, assigning signatures to all other components of $\piO$, 
we introduce  the norms,
\begin{equation}
\begin{split}
\PiO_0&=\ep^{-1}\|\piO\|_{\Lsc^4(S)}+\ep^{-1}\|\piO\|_{\Lsc^\infty(S)},\\
\PiO_1&=\sum_{(\mu,s)\ne (3,0)} \ep^{-1} \|D_\mu \piO^{(s)}\|_{\Lsc^4(S)}\\
&+
\ep^{-1} \|D_3\piO^{(0)}-\nab_3 Z\|_{\Lsc^4(S)}+\ep^{-1}
\|\sup_{\ub} |\nab_3 Z|\|_{\Lsc^2(S)},
\end{split}
\end{equation}
 \subsection{Proof of  Main  Theorem  I}
 To prove the theorem we start by making a bootstrap assumption
on the Ricci coefficient norm $\OO$. More precisely we assume
that,
\bea
\label{eq:bootstrap}
\OO\les \De_0
\eea
Based on this assumption   we  state 
various   preliminary    estimates  in section
 \ref{sect:preliminary}, which are simple adaptation 
  of results proved in \cite{K-R:trapped}. It is 
  interesting to remark that this is the only place
  when we need to make  a restriction for 
 the size of   $\de^{1/2}\ep^{-1}$. 
   Using these preliminary estimates we  then indicate 
    how, by a simple adjustment of the curvature estimates
    in  \cite{K-R:trapped} we can prove, see section \ref{sect:global.curv},   the following.  
      \begin{theorem}[Theorem A]
      \label{theoremA}
There exists a positive constant $a>\frac 18$ such that,
for      $\de^{1/2}\ep^{-1}$ and $\ep$ sufficiently small,
\bea
\RR(u,\ub) + \RRb(u,\ub)\les \RR^{(0)}+ C \ep^a ( \RR+\RRb)
\eea
with $C=C(\Pi,\Pib,\PiO, \RR,\RRb)$.
\end{theorem}
Next we rely on a theorem which  bounds the norms
$\Pi, \Pib $ and $\PiO$, for the deformation tensors of $L, \Lb$ and $O$,  to  the Ricci coefficients norms
$\OO$. 
\begin{theorem}[Theorem B]
\label{theoremB}
 Under the assumptions $\de^{1/2}\ep^{-1}$  and $\ep$ sufficiently small
 we have, 
  \bea
 \Pi+\Pib+\PiO\les C(\OO, \RR, \RRb)
 \eea
 \end{theorem}
 Finally we state the theorem which relates  the norms $\OO$ to the curvature norms $\RR, \RRb$. 

\begin{theorem}[Theorem C]
      \label{theoremC}
   Under the assumptions $\de^{1/2}\ep^{-1}$  and $\ep$ sufficiently small
   we have, with a constant $C=C(\OO^{(0)}, \RR, \RRb)$,
   \bea
   \OO&\les&C(\OO^{(0)}, \RR, \RRb)\label{eq:finalOO}
   \eea
  \end{theorem}
  Combining theorems B and C with theorem A
  we deduce, under the bootstrap assumption  \ref{eq:bootstrap},
   \beaa
\RR(u,\ub) + \RRb(u,\ub)\les \RR^{(0)}+\ep^a C( \RR,\RRb)( \RR+\RRb),
\eeaa
  from which, for $\ep$ sufficiently small,
  \bea
  \RR(u,\ub) + \RRb(u,\ub)\les \RR^{(0)}.
  \eea
Thus, back to \eqref{eq:finalOO} and using also proposition
\ref{thm.main.initial},
\beaa
\OO&\les& C( \RR^{(0)})
\eeaa
which allows us to remove the bootstrap assumption and 
confirm the result of the main theorem I.

\section{Formation of pre-scars}    
Relying on the results of  theorem \ref{thm:main} we  prove a new result
concerning the formation of pre-scars.   Throughout this section we assume that the assumptions and conclusions of theorem \ref{thm:main} hold true.

\subsection{Local scale invariant norms} 
Consider a partition of $S_0=S(0,0)$ into 
angular sectors  $\La$ of a given  size $|\La|$. 
Let $\fLa_{(0)}$ be a partition of unity associated 
to this partition, They can be    extend trivially, 
 first along $\Hb_0$ and then along  each  $H_u$, 
   to be constant along  the corresponding  null 
   generators. In particular we have,
   \bea
   \nab_L \fLa=0, \qquad \fLa|_{\, \Hb_0}=\fLa_{(0)}
   \eea
Then,  under the assumptions and conclusions of theorem \ref{thm:main}
we can easily deduce,
\begin{lemma}  
\label{le: fLa}
  We have,
\bea
\sum_{\La}\fLa=1
\eea
Also,
\bea
| \nab\fLa|_{L^{\infty}   }\les    |\La|^{-1},\qquad    | \nab_\Lb\fLa|_{L^{\infty}}\les\ep\de^{1/2}  |\La|^{-1}\label{eq:fLa}
\eea
or, in scale invariant norms (assigning to $f$ signature $0$), 
\beaa
| \nab\fLa|_{\Lsc^{\infty}   }\les  \de^{1/2} |\La|^{-1},\qquad  | \nab_\Lb\fLa|_{\Lsc^{\infty}}\les\ep  \de^{1/2}  |\La|^{-1}
\eeaa
\end{lemma}
We now   introduce the    localized  curvature  norms,
 \begin{equation}
 \begin{split}
  \RRLa_0(u,\ub):&=\|\fLa\a\|_{\Lsc^2(H_u^{(0,\ub)})}+\RRLa_0'(u,\ub')\\
 \RRLa_0'(u,\ub):&=   \ep^{-1} \|  \fLa(\b,\rho,\si,\bb,K)\|_{\Lsc^2(H_u^{(0,\ub)})}\\
   \RRLa_1(u,\ub):&= \|\fLa\nab_4\a\|_{\Lsc^2(H_u^{(0,\ub)})}+\RRLa_1'(u,\ub)\\
\RRLa_1'(u,\ub):&=    \ep^{-1}\|\fLa\nab( \a, \b,\rho,\si, \bb,K)\|_{\Lsc^2(H_u^{(0,\ub))})}\\
   \\
    \RRbLa_0(u,\ub):&=\|\fLa\b\|_{\Lsc^2(\Hb_{\ub}^{(u,0)})}+\RRb'_0(u,\ub')\\
  \RRbLa'_0(u,\ub)&=  \ep^{-1}\|\fLa(\rho,\si,\bb,\aa,K)\|_{\Lsc^2(\Hb_{\ub}^{(0,u)})}\\
   \RRbLa_1(u,\ub):&= \|\fLa\nab_3\aa\|_{\Lsc^2(\Hb_{\ub}^{(u,0)})}+ \RRb'_1(u,\ub)\\
   \RRb'_1(u,\ub): &=   \ep^{-1}      \|\fLa\nab(  \b,\rho,\si, \bb,\aa,K)\|_{\Lsc^2(\Hb_{\ub}^{(0,u)})}
   \end{split}
    \end{equation}
    and,
      \begin{equation}
 \begin{split}
  \RRLasup_0(u,\ub):&=\sup_{\La}  \RRLa_0,\qquad 
   \RRLasup_1(u,\ub):=\sup_{\La}  \RRLa_1\\
    \RRbLasup_0(u,\ub):&=\sup_{\La}  \RRbLa_0,\qquad 
   \RRbLasup_1(u,\ub):=\sup_{\La}  \RRbLa_1
  \end{split}
    \end{equation}
     with the supremum taken with respect to all elements of the partition.
    and,
    \bea
     \RRLasup= \RRLasup_0+ \RRLasup_1,\qquad  \RRbLasup= \RRbLasup_0+ \RRbLasup_1
    \eea
   
   \subsection{  Angular localized curvature estimates }
   Using a variation of our main energy estimates, with an additional angular    
localization, we can prove the following.

 \begin{theorem}  \label{thm.main.loc} 
 Under the assumptions and conclusions of theorem \ref{thm:main},
  if in addition $\de^{\frac 12} |\La|^{-1}$ is sufficiently small,  then,  for  $0\le u\le 1$, $0\le\ub\le \de$,
 \beaa
 (\RRLasup+\RRbLasup)(u,\ub)\les \RRLasup^{(0)}
 \eeaa
Moreover,
\bea
 (\RRLa+\RRbLa)(u,\ub)\les \RRLa^{(0)}+\de^{\frac 12} |\La|^{-1}\RRLasup^{(0)}\label{eq:spill}
 \eea
 \end{theorem}
  \begin{remark}
 By the standard domain of dependence argument the energy estimate can not fully localized to 
 individual sectors $\,^{(\La)} H_{u}$ and $\,^{(\La)} \Hb_{\ub}$ contained in the support of the function 
 $\,^{(\La)} f$. This explains the need for the 
 supremum in $\La$ in the definition  of the $\RRLasup, \RRbLasup$ norms for the first part of the 
 theorem. The second part of the theorem gives a bound for each  sector individual $\La$ with the second term
 on the right hand side of \eqref{eq:spill} accounting for the defect of localization.
 \end{remark}
A proof of the  theorem is sketched  in section \ref{sect:loc.curv}. 

\subsection{Emerging scars}
 \begin{definition}
We say that the data $\RR^{(0)}$ is uniformly distributed on the scale $\de^{\frac 12}\varpi^{-1}$  if there exists a 
partition $\{\La\}$ such that $|\La|\approx \de^{\frac 12} \varpi^{-1}$ and 
\bea
\RRLasup^{(0)}\les \de^{\frac 12}\varpi^{-1} \RR^{(0)}
\eea
\end{definition}
Our second main result of this paper is the following.
\begin{theorem}[Main theorem II] 
\label{thm:mainII}
Assume that, in additions to 
 the conditions  of validity of  theorem \ref{thm:main}, 
 the data $\RR^{(0)}$ is  uniformly distributed on
  the scale $\de^{\frac 12} \varpi^{-1}$ for some 
constant $\varpi<<1$ and 
$\ep\varpi^{-1}$ sufficiently small. Let $\La$ be a fixed  angular sector 
 of size $|\La|=q^{-1} \de^{\frac 12}$ with  $q=\ep\varpi^{-1}$
 sufficiently small.    Then, if  
\bea
\inf_{\th\in \La} \int_0^\de |\chih_0|^2(\ub,\th) d\ub >\frac {2(r_0-u)}{r_0^2}
\eea
 the $\La$-angular section  $\,^{(\La)} S_{u,\de}$ of the surface $S_{u,\de}$
must be  trapped, i.e. $\trch<0$ there.

 Alternatively, if for some constant $c>0$ independent of $\de,\ep,q,\varpi$,
 \bea
\sup_{\th\in \La} \int_0^\de |\chih_0|^2(\ub,\th) d\ub<\frac {2(r_0-u)}{r_0^2}-c
\eea
then  $\trch >0$ throughout   the angular sector   $\,^{(\La)} S_{u,\de}$.
\end{theorem}
We postpone a discussion of  the proof of this  theorem to the last section 
of the paper.
\begin{remark} 
Observe that the  parameters  $\de, \ep, \varpi$ in theorem \ref{thm:mainII}
verify the conditions:
\beaa
0<\de^{1/2} < \ep< \varpi< 1, \qquad     \de^{1/2}\ep^{-1}<< 1, \quad  q=\ep\varpi^{-1}<< 1.
\eeaa

\end{remark}

 \section{Preliminary estimates}
 \label{sect:preliminary}
  \subsection{Transported coordinates}
  As mentioned in the previous section 
we define systems of, local, transported coordinates along  the null 
 hypersurfaces $H$ and $\Hb$.  Staring with a local coordinate system $\th=(\th^1, \th^2)$
on   $U\subset  S(u,0)\subset H_u$ we parametrize any point along the null geodesics starting
 in $ U$ by the  the corresponding coordinate $\th$ and affine parameter $\ub$. Similarly,
 starting with a local coordinate system $\thb=(\thb ^1, \thb^2)$  on 
 $V\subset S(0,\ub)\subset \Hb_{\ub}$
 we parametrize  any point   along the null geodesics starting
 in $ V$ by the  the corresponding coordinate $\thb$ and affine parameter $u$.
 We denote 
the respective metric components by $\gamma_{ab}$ and  $\underline\gamma_{ab}$. 
\begin{proposition}\label{prop:gamma}
Let $\gamma^0_{ab}$ denote the standard metric on ${\Bbb S}^2$. Then, 
for any $0\le u\le 1$ and $0\le\ub\le\de$ and sufficiently small $\de^{\frac 12}\Delta_0$

$$
|\gamma_{ab}-\gamma^0_{ab}|\le \de^{\frac 12} \Delta_0,\qquad 
|\underline\gamma_{ab}-\gamma^0_{ab}|\le \de^{\frac 12} \Delta_0.
$$
The Christoffel symbols $\Gamma_{abc}$ and $\Gab_{ab}$,  obey the scale invariant  estimates\footnote{We  attach signature  $1/2$ to both  $\Gamma$
and $\Gab$.}

\begin{align}
&\|\Gamma_{abc}\|_{\Lsc^2(S)}\les \ep\OS_{[1]},\qquad \|\pr_d \Gamma_{abc}\|_{\Lsc^2(S)}\les\ep \OS_{[2]},\label{eq:crist-sc}\\
&\|\Gab_{abc}\|_{\Lsc^2(S)}\les \ep \OS_{[1]},\qquad \|\pr_d \Gab_{abc}\|_{\Lsc^2(S)}\les\ep\OS_{[2]},\label{eq:cristb-sc}
\end{align}
\end{proposition}
The proof is a trivial  adaptation of proposition 4.6 in \cite{K-R:trapped}.
 \subsection{Calculus inequalities}
\label{subs.calculus.ineq}
We simply adapt here the results  of  section 4.9  in \cite{K-R:trapped}.

\begin{proposition}\label{cor:interpol}
Let $S=S_{u,\ub}$ and let $S_\ep\subset S$ denote a disk of radius $\ep$ relative to either 
$\theta$ or $\underline\theta$ coordinate system. Then for any horizontal tensor $\phi$ and any $p>2$
\begin{align}
&\|\phi\|_{\Lsc^4(S)}\les \|\psi\|_{\Lsc^2(S)}^{\frac 12}  \|\nab\phi\|_{\Lsc^2(S)}^{\frac 12} +
 \de^{\frac 14}\|\phi\|_{\Lsc^2(S)},\label{eq:L4-glob-sc}\\
&\|\phi\|_{\Lsc^\infty(S)}\les \|\psi\|_{\Lsc^p(S)}^{\frac p{p+4}}  \|\nab\phi\|_{\Lsc^p(S)}^{\frac 4{p+4}} + 
 \de^{\frac 1p}\|\phi\|_{\Lsc^p(S)}.\label{eq:Linf-glob-sc}
\end{align}
and 
\begin{align}
&\|\phi\|_{\Lsc^4(S_\ep)}\les \|\nab\phi\|_{\Lsc^2(S_{2\ep})}+
(\ep^{-2} \de)^{\frac 14} \|\phi\|_{\Lsc^2(S_{2\ep})},\label{eq:L4-loc-sc}\\
&\|\phi\|_{\Lsc^\infty(S)}\les \sup_{S_\ep\subset S} \left (\|\nab\phi\|_{\Lsc^4(S_{2\ep})}+
(\ep^{-2} \de)^{\frac 14}\|\phi\|_{\Lsc^4(S_{2\ep})}\right).\label{eq:Linf-loc-sc}
\end{align}
\end{proposition}
As a consequence of the proposition we derive.
\begin{corollary}
$$
\OS_{0,\infty}\les  \OS_{1,2}^{\frac 12}\c \OS_{2,2}^{\frac 12}+(\ep^{-2}\de)^{\frac 14} \OS^\ep_{0,4}
$$
\end{corollary}  
\subsection{ Codimension $1$ trace formulas.}
The following is a straightforward  adaptation of proposition 4.15 
in \cite{K-R:trapped}
 \begin{proposition}
 \label{prop.trace.sc}
 The following formulas hold true for  a fixed  $S=S(u,\ub)=H(u)\cap \Hb(\ub)\subset\DD $ and
any horizontal tensor $\phi$
\beaa
\|\phi\|_{\Lsc^4(S)}&\les&\big(\de^{1/2} \|\phi\|_{\Lsc^2(H)}+\|\nab\phi\|_{\Lsc^2(H)}\big)^{1/2}\big(\de^{1/2}\|\phi\|_{\Lsc^2(H)}+\|\nab_4\phi\|_{\Lsc^2(H)}\big)^{1/2}\\
\|\phi\|_{\Lsc^4(S)}&\les&\big(\de^{1/2}\|\phi\|_{\Lsc^2(\Hb)}+\|\nab\phi\|_{\Lsc^2(\Hb)}\big)^{1/2}\big(\de^{1/2} \|\phi\|_{\Lsc^2(\Hb)}+\|\nab_3\phi\|_{\Lsc^2(\Hb)}\big)^{1/2}\
\eeaa
 \end{proposition}
 \subsection{Estimates for  Hodge systems}
 \label{section.Hodge}
 Here we make straightforward adaptations of the results (more precisely  propositions 4.17 and 4.17)  in section
 4.16 of \cite{K-R:trapped} for Hodge systems.
\begin{proposition}
Let $\psi$ verify the Hodge system
\bea
\DD \psi=F,
\eea
with $\Dcal$ 
one of the Hodge  operators
defined   in section 3.5    of \cite{K-R:trapped}.
Then,
\bea
\|\nab\psi\|_{\Lsc^2(S)}&\les&\|K\|_{\Lsc^2(S)} \|\psi\|_{\Lsc^2(S)}+\|F\|_{\Lsc^2(S)}
\eea
\label{prop:Hodge.estim}
\end{proposition}
Also,
\begin{proposition}
Let $\psi$ verify the Hodge system
\bea
\DD \psi=F
\eea
Then,
\bea
\|\nab^2\psi\|_{\Lsc^2(S)}\les\|K\|_{\Lsc^2(S)} \|\psi\|_{\Lsc^\infty(S)}+ \|K\|^{\frac 12}_{\Lsc^2(S)} 
\|\nab\psi\|_{\Lsc^4(S)}
+\|\nab F\|_{\Lsc^2(S)}
\eea
\label{prop:Hodge.estim-2}
\end{proposition}

\subsection{Trace theorems} 
We state the straightforward adaptations of the results of section 11
in \cite{K-R:trapped} concerning sharp trace theorems.

We introduce the following trace   norms for an $S$ tangent tensor $\phi$, with signature  $\sgn(\phi)$,  along
 $H=H_u^{(0,\ub)}$, relative to the transported coordinates $(\ub, \th)$ of proposition \ref{prop:gamma}:
\beaa
 \|\phi\|_{\Trsc(H)}&=&\de^{\sgn(\phi) - \frac 1 2 } \big( \,\sup_{\th\in S(u,0)} \int_0^{\ub}|\phi(u,\ub', \th)|^2 d\ub' \big)^{1/2}
 \eeaa
 Also,  along $\Hb=\Hb_{\ub}^{(0,u)} $
  relative to the transported coordinates $(u, \thb)$ of proposition \ref{prop:gamma}
  \beaa
\|\phi\|_{\Trsc(\Hb)}&=&\de^{\sgn(\phi)  } \big(\, \sup_{\thb\in S(\ub,0)} \int_0^{u}|\phi(u',\ub,\thb)|^2 d u' \big)^{1/2} \eeaa
\begin{proposition}
\label{prop.trace}
For any horizontal tensor $\phi$ along $H=H_u^{(0,\ub)}$,
\begin{equation}
\label{eq:*}
\begin{split}
\|\nab_4\phi\|_{\Tr_{(sc)}({H})}&\les \left (\|\nab_4^2\phi\|_{\Lsc^2(H)}+\|\phi\|_{\Lsc^2(H)}+ 
\ep C(\|\phi\|_{\Lsc^\infty}+
\|\nab_4\phi\|_{\Lsc^4(S)})\right)^{\frac 12}\\ &\times  \left (\|\nab^2\phi\|_{\Lsc^2(H)}+ \ep C(\|\phi\|_{\Lsc^\infty}+
\|\nab\phi\|_{\Lsc^4(S)})\right)^{\frac 12}\\ &+\|\nab_4\nab\phi\|_{\Lsc^2(H)}+ \|\phi\|_{\Lsc^\infty}+\|\nab\phi\|_{\Lsc^2(H)}
\end{split}
\end{equation}
 where
 $C$ is a constant which 
depends  on $\OO^{(0)}, \RR, \RRb$.

Also, 
for any horizontal tensor $\phi$ along $\Hb=H_{\ub}^{(u,0)}$, and
a similar constant $C$,
\begin{equation}
\label{eq:*b}
\begin{split}
\|\nab_3\phi\|_{\Tr_{(sc)}({\Hb})}&\les \left (\|\nab_3^2\phi\|_{\Lsc^2(\Hb)}+\|\phi\|_{\Lsc^2(\Hb)}+ 
\ep C(\|\phi\|_{\Lsc^\infty}+
\|\nab_3\phi\|_{\Lsc^4(S)})\right)^{\frac 12}\\ &\times  \left (\|\nab^2\phi\|_{\Lsc^2(H)}+ \ep C(\|\phi\|_{\Lsc^\infty}+
\|\nab\phi\|_{\Lsc^4(S)})\right)^{\frac 12}\\ &+\|\nab_3\nab\phi\|_{\Lsc^2(\Hb)}+\|\phi\|_{\Lsc^\infty}+
\|\nab\phi\|_{\Lsc^2(\Hb)}
\end{split}
\end{equation}

\end{proposition}

   \section{ Global Curvature Estimates }
   \label{sect:global.curv}
   In this section we  discuss the proof of theorem A, \ref{theoremA},
   which is a straightforward modification of the curvature estimates of
   sections 14 and 15 in \cite{K-R:trapped}.

  \subsection{Zero order estimates} 
  As in \cite{K-R:trapped} all  curvature estimates are based on the energy identities for the Bel-Robinson tensor $\QQ[W]$
of a  Weyl field $W$ which we take  here to be either  
  the Riemann curvature tensor $R$  or its modified 
 Lie derivatives $\Lieh_U R=\Lie_U R -\frac 1 8 \tr  ^U\pi R-    \frac 1 2 ^U\pih\c R $, relative to well chosen vectorfields $U$. Recall 
 \begin{proposition} 
 \label{cor;0energy}
 The following identity holds on our fundamental domain $\DD(u,\ub)$,
 \beaa
 \int_{H_u} Q[R] (L, X, Y, Z)+\int_{\Hb_{\ub}} Q[R] ( X, Y, Z, \Lb)
&=&\int_{H_0}Q[R] (L, X, Y, Z)\\
&+&
\frac 1 2 \int\int_{\DD(u, \ub)}Q[R]\c \pi(X,Y, Z),
\eeaa
where $\pi(X,Y,Z)$ is a linear combination of the deformation tensors of the vectorfields $X, Y, Z$. 
\end{proposition}
The global  estimates  corresponding to the norms $\RR_0$ and $\RRb_0$
 are obtained, as in section 14 of  \cite{K-R:trapped} by making  the choices 
$
(X,Y,Z)=\{(L,L,L); (L,L,\Lb); (L,\Lb,\Lb); (\Lb,\Lb,\Lb)\}.
$
and following precisely the same steps as before.
We summarize the result in  the following,
\begin{proposition}
\label{prop:curvzero}
There exists a positive constant $a>\frac 18$ such that,
for      $\de^{1/2}\ep^{-1}$ and $\ep$ sufficiently small,
\bea
\RR_0(u,\ub)+\RRb_0(u,\ub)\le \RR_0(0,\ub)+\ep^a C(\Pi_0,\Pib_0,\RR,\RRb)(\RR+\RRb)
\eea
\end{proposition}
   We sketch the proof in the particular case when  
    $X=Y=Z=L$ in proposition \ref{cor;0energy}.  We obtain, schematically, by signature considerations, 
\beaa
 \int_{H_u} |\a|^2+\int_{\Hb_{\ub}} |\b|^2
&=&\int_{H_0}|\a|^2
+\frac 3 2 \int\int_{\DD(u, \ub)}Q[R](\piL,L,L)\\
&\les&\int_{H_0}|\a|^2+\sum_{s_1+s_2+s_3=4}
 \piL^{(s_1)}\c \Psi^{(s_2)}\c \Psi^{(s_3)}
\eeaa
Passing to the scale invariant norms we have,   
\begin{align*}
\|\a\|^2_{\Lsc^2(H_u^{({0,\ub})}}+ \|\b\|^2_{\Lsc^2(\Hb_{\ub}^{({0,u})}}
&\le \|\a\|^2_{\Lsc^2(H_0^{({0,\ub})}}+
\sum_{s_1+s_2+s_3=2s=4} \de^{2}\int\int_{\DD(u, \ub)} \piL^{(s_1)}\c \Psi^{(s_2)}\c \Psi^{(s_3)}
\end{align*}
 The worst term occur when $s_2=s_3=2$ and $s_1=0$. 
  Observe also that, since the signature
of a Ricci coefficient $\piL^{(s_1)}$ may not exceed $s_1=1$, neither 
$s_2$ or $s_3$ can be zero, i.e. $\aa$ cannot occur  among the curvature terms
 on the right.   We use the estimate $   \|\piL^{(s_1)}\|_{\Lsc^\infty} \le  \ep  \Pi_0$ to  deduce,
 \beaa
\|\a\|_{\Lsc^2(H^{(0,\ub)}_u)}^2 +\|\b\|_{\Lsc^2(H^{(0,u)}_{\ub})}^2 
&\les& \|\a\|_{\Lsc^2(H^{(0,\ub)}_0)}^2+ \ep \Pi_0 \c \RR_0(u,\ub)^2
\eeaa
There other estimates  are derived in the same manner, see \cite{K-R:trapped}

\subsection{First derivative estimates}As in \cite{K-R:trapped}
the first derivative curvature estimates are based on the  following.
\begin{proposition}
\label{corr:mainenergy1} Let $U$ be a vectorfield  defined  in   our fundamental domain $\DD(u,\ub)$,  tangent to $\Hb_0$. Then,  with $H_u=H_u([0,\ub])$,
 \beaa
&& \int_{H_u} Q[\Lieh_U R] (L, X, Y, Z)+\int_{\Hb_{\ub} }Q[\Lieh_UR](  X, Y, Z,\Lb)=
\int_{H_0}Q[\Lieh_UR] (L, X, Y, Z)\\
&&+\frac 1 2 \int\int_{\DD(u, \ub)}Q[\Lieh_U R]\c \pi(X,Y, Z)+\int\int_{\DD(u, \ub)}D(R, U)(X,Y, Z)
\eeaa
with $ D(U, R)$ the $3$-tensor  of the form, schematically.
\beaa
 D(U, R)&=(\Lieh_U R)\c\big(\pi\c D R+D\pi \c R)
\eeaa
\end{proposition}
We apply these  estimate for the following  the choice of vectorfields,
\begin{align*}
(U; X,Y,Z)=\{(L; L,L,L); (\Lb; \Lb,\Lb,\Lb); (O; L,L,L); (O; L,L,\Lb); (O; L,\Lb,\Lb); (O; \Lb,\Lb,\Lb)\},
\end{align*}
As in \cite{K-R:trapped}, see section 15,  we make 
the choice $(U; X,Y,Z)=(L; L,L,L)$ to  the estimate  $\nab_4\a$ and the choice   $(U; X,Y,Z)=(\Lb; \Lb,\Lb,\Lb)$  to estimate
 $\nab_3\alphab$.    The four  choices   $U=O $ and $X, Y, Z\in\{L, \Lb\}$ lead to  bounds for 
$\nab\a, \nab\b, \nab(\rho,\si), \nab\bb$, which coupled with the Bianchi identities are sufficient to estimate 
all first derivatives of the null curvature components.  
We outline below   a typical estimate involving $O$. 
 Let $\Psi^{(s)}(\Lieh_O R)$ and 
$\Psi^{(s)}(D R)$ denote the null components of
the Weyl field  $\Lieh_O R$ and $DR$ of signature $s$. Then
\begin{align*}
 \int_{H_u} |\Psi^{(s)}(\Lieh_O R)|^2 &+  \int_{\Hb_{\ub}} |\Psi^{(s-\frac 12)}(\Lieh_O R)|^2=
 \int_{H_0} |\Psi^{(s)}(\Lieh_O R)|^2 \\ &+ \sum_{s_1+s_2+s_3=2s}\int\int_{\DD(u, \ub)} 
 (\piL^{(s_1)},\piLb^{(s_1)})\c \Psi^{(s_2)}(\Lieh_O R)\c \Psi^{(s_3)}(\Lieh_O R)\\ &+
 \sum_{s_1+s_2+s_3=2s}\int\int_{\DD(u, \ub)} 
 \piO^{(s_1)}\c \Psi^{(s_2)}(D R)\c \Psi^{(s_3)}(\Lieh_O R)\\ &+
 \sum_{s_1+s_2+s_3=2s}\int\int_{\DD(u, \ub)} 
 (D\piO)^{(s_1)}\c \Psi^{(s_2)}\c \Psi^{(s_3)}(\Lieh_O R)
\end{align*}
Using our scale invariant norms, and proceeding exactly as 
in section 15 of \cite{K-R:trapped} 
we can easily derive the estimate, for some $a>\frac 1 8$,
\begin{align*}
\sum_{s\in \{\frac 12,1,\frac 32,2\}} 
\ep^{-1}
\left(\|\nab\Psi^s\|^2_{\Lsc^2(H_u^{({0,\ub})}}+\|\nab\Psi^{s-\frac 12}\|^2_{\Lsc^2(\Hb_{\ub}^{({0,u})}}\right)
&\les \sum_{s\in \{\frac 12,1,\frac 32,2\}} 
\ep^{-1}\|\nab\Psi^s\|^2_{\Lsc^2(H_0^{({0,\ub})}}
\\ &+\ep^a C(\Pi_0,\Pib_0,\PiO_0, \RR,\RRb)(\RR+\RRb),\\
\end{align*}
Similarly, 
\begin{align*}
&\|\nab_4\a\|^2_{\Lsc^2(H_u^{({0,\ub})}}\les 
\|\nab_4\a\|^2_{\Lsc^2(H_0^{({0,\ub})}}+\ep^a C(\Pi_0,\Pib_0,\Pi_1,\RR,\RRb)(\RR+\RRb),\\
&\|\nab_3\alphab\|^2_{\Lsc^2(\Hb_{\ub}^{({0,u})}}
\les \|\nab_3\aa\|^2_{\Lsc^2(\Hb_0^{({0,u})}}
+\ep^a C(\Pi_0,\Pib_0,\Pib_1,\RR,\RRb)(\RR+\RRb),
\end{align*}

Combining,  we derive the desired first derivative estimates 
\begin{proposition}
\label{prop:deriv.estim}
There exists a positive constant $a>\frac 18$ such that,
for      $\de^{1/2}\ep^{-1}$ and $\ep$ sufficiently small,
$$
\RR_1(u,\ub) + \RRb_1(u,\ub)\les \RR_1(0,\ub) + C \ep^a(\RR+\RRb)
$$
with $C= C(\Pi,\Pib,\PiO, \RR,\RRb)$.
\end{proposition}
Combining this with proposition \ref{prop:curvzero} we derive,
\bea
\RR(u,\ub) + \RRb(u,\ub)\les \RR^{(0)}+ C\ep^a (\RR+\RRb)
\eea
which ends the proof  of  theorem \ref{theoremA}.

\section{Localized Energy estimates}
\label{sect:loc.curv}
\subsection{Localized  zero order estimates}
We start by modifying slightly proposition \ref{cor;0energy},
 \begin{proposition} 
 \label{cor;0energy}
 The following identity holds on our fundamental domain $\DD(u,\ub)$,
 \beaa
&& \int_{H_u} \fLa^2 Q[R] (L, X, Y, Z)+\int_{\Hb_{\ub}} \fLa^2Q[R] ( X, Y, Z, \Lb)
=\int_{H_0} \fLa^2Q[R] (L, X, Y, Z)\\
&&+
\frac 1 2 \int\int_{\DD(u, \ub)} \fLa^2Q[R]\c \pi(X,Y, Z)
+2 \int\int_{\DD(u, \ub)} \fLa Q[R](D\fLa, X, Y, Z)
\eeaa
where $\pi(X,Y,Z)$ is a linear combination of the deformation tensors of the vectorfields $X, Y, Z$. 
\end{proposition}
As in the derivation of the global estimates we make all   the choices, 
\beaa
(X,Y,Z)=\{(L,L,L); (L,L,\Lb); (L,\Lb,\Lb); (\Lb,\Lb,\Lb)\}.
\eeaa
In each case the only new term that needs to be estimated
is due to $ \int\int_{\DD(u, \ub)} \fLa Q[R](D\fLa, X, Y, Z)$. 
Consider again the particular case  $X=Y=Z=L$. Then,
\beaa
 \int_{H_u} |\fLa \, \a|^2+\int_{\Hb_{\ub}} |\fLa \, \b|^2
&=&\int_{H_0}|\fLa \, \a|^2
+\frac 3 2 \int\int_{\DD(u, \ub)} \fLa^2 \, Q[R](\piL,L,L)\\
&+&2 \int\int_{\DD(u, \ub)}Q[R](D\fLa, L, L, L)
\eeaa
Clearly, recalling \eqref{eq:fLa},
\beaa
 |Q[R](D\fLa, L, L, L)|&\les& | \nab_3\fLa|  |\a|^2+ | \nab\fLa|   |\b|\c|\a|\\
 &\les&\ep  \de^{1/2}  |\La|^{-1}  |\a|^2+   |\La|^{-1}   |\b|\c|\a|
\eeaa
Recalling also,  $\sum_{\Lat} \fLat=1$, and  $\fLa\c\fLat=0$  except for a
a few neighboring $\Lat$, 
\beaa
  | \fLa Q[R](D\fLa, L, L, L)|&\les&\ep  \de^{1/2}  |\La|^{-1}    \sum_{\Lat}     |\fLa\, \a| \c |\fLat\, \a|   + |\La|^{-1}    |\fLa\, \a|\c|\fLat\, \b|\\
  &\les&\ep  \de^{1/2}  |\La|^{-1}  |\fLa\, \a| \c\sup_{\Lat}  |\fLat\, \a| 
  + |\La|^{-1}    |\fLa\, \a|\c \sup_{\Lat}|\fLat\, \b|
  \eeaa
Therefore,  passing to scale invariant norms, and treating the term
in $\piL$  exactly as before,
\beaa
\|\fLa\,\a\|_{\Lsc^2(H^{(0,\ub)}_u)}^2 +\|\fLa\,\b\|_{\Lsc^2(H^{(0,u)}_{\ub})}^2 &\les& \|\fLa\a\|_{\Lsc^2(H^{(0,\ub)}_0)}^2     +      \ep  \Pi_0\c  \RRLa_0^2(u,\ub)\\
&+&  \ep \de^{1/2} |\La|^{-1}   \| \fLa\, \a\|_{\Lsc^2(H_u^{(0,\ub)})} \c \sup_{\Lat} \| \fLat\, \a\|_{\Lsc^2(H_u^{(0,\ub)})}\\
&+&
  |\La|^{-1} \|\fLa\, \a\|_{\Lsc^2(H_u^{(0,\ub)})} \c \de^{1/2}   \sup_{\Lat}   \|\fLat\,  \b\|_{\Lsc^2(H_u^{(0,\ub)})}\\
 &\les& \|\fLa\a\|_{\Lsc^2(H^{(0,\ub)}_0)}^2     +      \ep  \Pi_0\, \RRLa_0^2(u,\ub)\\
 &+& |\La|^{-1}\de^{1/2} \big(\ep  \RRLa_0[\a]\c\RRLasup_0[\a]+ \RRLa_0[\a]\c\RRLasup_0[\b]\big)
\eeaa
Therefore, taking the supremum over $\La$ on  both sides, we derive
\beaa
\RRLasup^2_0[\a](u,\ub)+\RRbLasup^2_0[\b](u,\ub)&\les&\RRLasup^2_0[\a](0,\ub)
+\ep\,  \Pi_0\c \RRLasup_0^2(u,\ub)\\
&+& |\La|^{-1}\de^{1/2}\big(\ep\, \RRLasup^2_0[\a]+ \RRLasup_0[\a]\c \RRLasup_0[\b]\big)
\eeaa
Proceeding in the same manner with all other curvature components
we derive,
\begin{proposition}
\label{prop:curvzero-La}
Consider  a partition of unity  $\fLa$ of of length $|\La|$
such that $\de^{1/2}|\La|^{-1}$ is sufficiently small. 
There exists a positive constant $a>\frac 18$ such that,
for      $\de^{1/2}\ep^{-1}$ and $\ep$ sufficiently small,
\bea
\RRLasup_0(u,\ub)+\RRbLasup_0(u,\ub)\le \RRLasup_0(0,\ub)+\ep^a C(\Pi_0,\Pib_0,\RRLasup,\RRbLasup)
(\RRLasup+\RRbLasup)
\eea
\end{proposition}
\subsection{ Localized derivative estimates}
We start  with a localized version of proposition \ref{corr:mainenergy1}.
\begin{proposition}
\label{corr:mainenergy1-La} Let $U$ be a vectorfield  defined  in   our fundamental domain $\DD(u,\ub)$,  tangent to $\Hb_0$. Then,  with $H_u=H_u([0,\ub])$,
 \beaa
&& \int_{H_u}  \fLa^2Q[\Lieh_U R] (L, X, Y, Z)+\int_{\Hb_{\ub} }\fLa^2Q[\Lieh_UR](  X, Y, Z,\Lb)=
\int_{H_0}\fLa^2 Q[\Lieh_UR] (L, X, Y, Z)\\
&&+\frac 1 2 \int\int_{\DD(u, \ub)} \fLa^2 Q[\Lieh_U R]\c \pi(X,Y, Z)+\int\int_{\DD(u, \ub)}\fLa^2D(R, U)(X,Y, Z)\\
&&+2 \int\int_{\DD(u, \ub)} \fLa Q[\Lieh_U R](D\fLa, X, Y, Z)
\eeaa
with $ D(U, R)=(\Lieh_U R)\c\big(\pi\c D R+D\pi \c R)$. 
\end{proposition}
We apply these  estimate, as before,  for same   choice of vectorfields,
\begin{align*}
(U; X,Y,Z)=\{(L; L,L,L); (\Lb; \Lb,\Lb,\Lb); (O; L,L,L); (O; L,L,\Lb); (O; L,\Lb,\Lb); (O; \Lb,\Lb,\Lb)\},
\end{align*}
The only new terms which  need to be treated   are due to
 $ \int\int_{\DD(u, \ub)} \fLa Q[\Lieh_U R](D\fLa, X, Y, Z)$.
For all choices of vectorfields $U, ,X, Y, Z$ we can proceed 
precisely as in the proof of proposition \ref{corr:mainenergy1}
and thus derive.
\begin{proposition}
\label{prop:deriv.estim-La}
Given   a partition of unity  $\fLa$  of length $|\La|$,
such that $\de^{1/2}|\La|^{-1}$ is sufficiently small, we can find 
 $a>\frac 18$ such that, for      $\de^{1/2}\ep^{-1}$ and $\ep$ sufficiently small,
\beaa
\RRLasup_1(u,\ub) + \RRbLasup_1(u,\ub)\les \RR_1(0,\ub) +\ep^a C(\Pi,\Pib,\PiO, \RRLasup,\RRbLasup)(\RRLasup+\RRbLasup)
\eeaa
\end{proposition}
Combining propositions \ref{prop:deriv.estim} and \ref{prop:deriv.estim-La}
we derive,
\begin{theorem}
Given   a partition of unity  $\fLa$  of length $|\La|$,
such that $\de^{1/2}|\La|^{-1}$ is sufficiently small, we can find 
 $a>\frac 18$ such that, for      $\de^{1/2}\ep^{-1}$ and $\ep$ sufficiently small,
we have,
\bea
\RRLasup(u,\ub) + \RRbLasup(u,\ub)\les \RRLasup^{(0)}+\ep^a C(\Pi,\Pib,\PiO, \RRLasup,\RRbLasup)(\RRLasup+\RRbLasup)
\eea
\end{theorem}
\section{Deformation tensor estimates}
In this section we sketch the proof of the estimates 
which relate  the norms $\Pi$ of the deformation tensors
for $L,\Lb$ and $O$ to the Ricci coefficient norms $\OO$, stated in theorem
\ref{theoremB}.
Throughout the section we assume that  $ \de^{1/2}\ep^{-1} $ and $\ep$
are sufficiently small. 
\subsection{Estimates for $\Pi$ and $\Pib$}
 The null components of  $\piL$ and $\piLb$
 are simply expressed in terms of null Ricci coefficients according to the lemma.
\begin{lemma}
   Below we list the components of $ \piL_{\a\b}$ and $\piLb_{\a\b}$.
\begin{equation}
\label{}
\begin{split}
& \piL_{44}=0,\quad \piL_{43}=0,\quad ^L\pi_{33}=-2\Omega^{-1} \omb,\\
&\piL_{4a}=0,\quad \piL_{3a}=\Omega^{-1} (\eta_a+\zeta_a)+\Omega^{-1}\nab_a \log \Omega,\quad\\
 &\piL_{ab}=\Omega^{-1} \chi_{ab}
\end{split}
\end{equation}
and,
\begin{equation}
\begin{split}
&\piLb_{33}=0,\quad \piLb_{43}=0,\quad \piLb_{33}=-2\Omega^{-1} \om,\\
&\piLb_{3a}=0,\quad \piLb_{4a}=\Omega^{-1} (\etab_a+\zeta_a)+\Omega^{-1}\nab_a \log \Omega,\quad\\
& \piLb_{ab}=\Omega^{-1} \chib_{ab}
\end{split}
\end{equation}

\end{lemma}
As a result we can easily  derive  the  estimates,
\begin{align*}
&\Pi_{0,4}\les \OS_{0,4},\qquad \Pi_{0,\infty}\les \OS_{0,\infty},\\
&\Pib_{0,4}\les \OS_{0,4},\qquad \Pib_{0,\infty}\les \OS_{0,\infty}
\end{align*}
Similarly we can estimate the first derivative norms,
\beaa
\Pi_1\les C(\OS_{1,4},\OS_0),\qquad \Pib_1\les C(\OS_{1,4},\OS_0).
\eeaa
These can be summarized in the following:
\begin{proposition} The following estimates hold true
 for the deformation tensors  $\piL$ and $\piLb$.
 \bea
 \Pi+\Pib &\les& C(\OO)
 \eea
\end{proposition}
which establishes half of theorem B (\ref{theoremB}).
\subsection{Estimates for $\PiO$}
Recall that the only non-vanishing components of $\piO$
are given by 
\begin{align*}
&\pi_{34}=-2(\eta+\etab)_a O_a,\\
&\pi_{ab}=\frac 12 (\nab_a O_b+\nab_b O_a):=H_{ab}^{(s)} =\frac 12(H_{ab}+H_{ba}),\\
&\pi_{3a}=\frac 12 (\nab_3 O_a-\chib_{ab} O_b):=\frac 12 Z_a.
\end{align*}
 The quantities $Z$ and $H$  verify the following transport equations, written schematically,
 \begin{equation}
 \label{eq:mainHZ} 
 \begin{split}
& \nab_4 Z=\nab(\eta+\etab)\c O+(\etab-\eta)\c H + \om Z+(\si+\rho)\c O + (\eta-\etab)\c(\eta+\etab)\c O,\\
  &\nab_4 H=\chi\c H +\b\c O+\nab\chi\c O +\chi\c\etab\c O
  \end{split}
 \end{equation}
 In view of equations \eqref{eq:mainHZ} we derive, by integration,
 \beaa
 \|Z\|_{\Lsc^\infty} &\les&\|\nab(\eta,\etab)\|_{\Trsc}+\|(\rho,\si)\|_{\Trsc}+ \|\psi\|_{\Lsc^\infty}(\|\psi\|_{\Lsc^\infty}+\|H\|_{\Lsc^\infty}+\|Z\|_{\Lsc^\infty})\\
 \eeaa
 Using the trace  estimates 
 for $(\eta, \etab)$ and $(\rho,\si)$ we derive,
  \beaa
\ep^{-1} \|Z\|_{\Lsc^\infty(S)} &\les&C+C(\|H\|_{\Lsc^\infty}+\|Z\|_{\Lsc^\infty})
\eeaa
with a constant $C=C(\II_0,\OS,\RR,\RRb)$.
Similarly,
\beaa
\ep^{-1} \|H\|_{\Lsc^\infty} &\les&\ep^{-1}\left (\|\nab\chih\|_{\Trsc}+\|\nab\trch\|_{\Lsc^\infty}+ \|\psi\|^2_{\Lsc^\infty}(\|\psi\|_{\Lsc^\infty}+\|H\|_{\Lsc^\infty})\right)\\&\les& C+C(C+\|H\|_{\Lsc^\infty}),
\eeaa
Following precisely  the same steps as section 13 of  in \cite{K-R:trapped} we derive,
\beaa
\ep^{-1}\|\piO\|_{\Lsc^4(S)} + \ep^{-1}\|\piO\|_{\Lsc^\infty(S)}+\ep^{-1}\|D\pi\|_{\Lsc^2(S)}\les C=C(\OS, \RR, \RRb).
\eeaa
Also all null components  of the derivatives  $D\piO$, with the exception of  $(D_3\piO)_{3a}$,  verify the estimates, 
\beaa
\ep^{-1}\|D\piO\|_{\Lsc^4(S)  } \les C
\eeaa
Moreover,
\beaa
\ep^{-1}\|(D_3\piO)_{3a}-\nab_3 Z\|_{\Lsc^4(S)} +  \ep^{-1}\|\sup_{\ub}|\nab_3 Z|\|_{\Lsc^2(S)}  &\les & C
\eeaa
Recalling the definition of the norms $\PiO$ we  deduce,

\begin{proposition}
\label{thmn:thmB}      The following estimates 
hold true     with a constant $C=C(\OO, \RR, \RRb)$,
\bea
\PiO&\les&C(\OO, \RR, \RRb).
\eea

\end{proposition}
This establishes the remaining part of theorem B(\ref{theoremB}).

 \section{Estimates for the Ricci coefficients}
 In this section we discuss the proof of theorem C(\ref{theoremC}).
 We make the point that the proof can be derived by a 
  straightforward 
 modification of the arguments in sections 5-10 of \cite{K-R:trapped}.

  Relying on the bootstrap assumption 
   the boot-strap assumption \eqref{eq:bootstrap}
 we first derive,  see section 4.1. in  \cite{K-R:trapped},
 \beaa
 \|\Om^{-1}- 2\|_{L^\infty(u,\ub) }&\les& \int_0^u \|\omb\|_{L^\infty(u',\ub)} du'
 \les \ep \, \OS_{0,\infty}[\omb]\les \ep\, \De_0.
 \eeaa
 Thus, if $\ep $ is sufficiently small  we deduce  that $|\Om-\frac 1 2| $ is small and therefore,
\beaa
\frac 1 4 \le \Om\le 4.
\eeaa
Using this  fact  we can deduce, as in section  4.1. of 
 \cite{K-R:trapped},
 \begin{proposition}
 \begin{equation}
\label{eq:transp.Hu.sc}
\begin{split}
\|\ \psi\|_{\Lsc^p(u, \ub)   }&\les \| \psi\|_{\Lsc^p(u,0 )}+\int_0^{\ub} \de^{-1} \|   \nab_4\psi   \|_{\Lsc^p(u,\ub'  ) }\,d\ub'\\
\|\ \psi\|_{\Lsc^p(u, \ub)   }&\les \| \psi\|_{\Lsc^2(0,\ub  )}+\int_0^{u}  \|   \nab_3\psi   \|_{\Lsc^p(u',\ub  ) } \, du'.
\end{split}
\end{equation}
 \end{proposition}

 \subsection{Estimates for  $\chi,\eta, \omb$ }
 The null Ricci coefficients  $\chi, \eta$ and $\omb$
verify transport equations of the form,
\bea
\nab_4 \psi^{(s)}=\sum_{s_1+s_2=s+1}\psi^{(s_1)}\c\psi^{(s_2)}+\Psi^{(s+1)}\label{eq:transp.symb4}
\eea
we have 
\beaa
\|\psi^{(s)}\|_{\Lsc^4(u,\ub)}&\les& \|\psi^{(s)}\|_{\Lsc^4(u,0)}+
\int_0^{\ub} \de^{-1}  \|    \nab_4\psi^{(s)}   \|_{\Lsc^4(u,\ub'  )}
\eeaa
To estimate $ \|    \nab_4\psi^{(s)}   \|_{\Lsc^4(u,\ub'  )}$ we make us 
of   the scale invariant  estimates 
\beaa
\|\phi\cdot\psi\|_{\Lsc^4(S)}\les \|\phi\|_{\Lsc^\infty(S)} \|\psi\|_{\Lsc^4(S)}
\eeaa
Hence,
\beaa
 \|    \nab_4\psi^{(s)}   \|_{\Lsc^4(S )}&\les& \| \Psi^{(s+1)}   \|_{\Lsc^4(S )}+\sum_{s_1+s_2=s+1}\|\psi^{(s_1)} \|_{\Lsc^\infty(S)}\|\psi^{(s_2)} \|_{\Lsc^4(S)}
 \eeaa
If all  scale invariant norms were small, i.e. $O(\ep)$, we 
 would proceed in a straightforward manner 
as follows,
\beaa
 \|    \nab_4\psi^{(s)}   \|_{\Lsc^4(S )}&\les& \| \Psi^{(s+1)}   \|_{\Lsc^4(S )} +
\ep^2\OS_{0, \infty}\c \OS'_{0, 4}\\
 &\les&\| \Psi^{(s+1)}   \|_{\Lsc^4(S )} + \ep^2\De_0\c \OS'_{0, 4}
 \eeaa
This in fact works for  $s<1$, i.e. for  $\omb$ and $\eta$.
In that case we  have, by integration,
 \beaa
 \|\psi^{(s)}\|_{\Lsc^4(u,\ub)}&\les& \|\psi^{(s)}\|_{\Lsc^4(u,0)}+
\int_0^{\ub} \de^{-1} \| \Psi^{(s+1)}   \|_{\Lsc^4(u,\ub' )} +\ep^2\De_0\c \OS'_{0,4}\\
&\les&\|\psi^{(s)}\|_{\Lsc^4(u,0)}+ \| \Psi^{(s+1)}   \|_{\Lsc(H_u)}^{1/2}  \|\nab \Psi^{(s+1)}   \|_{\Lsc(H_u)}^{1/2} +\ep^2\De_0\c \OS'_{0,4}\\
&\les&\|\psi^{(s)}\|_{\Lsc^4(u,0)}+\ep (\RR'_0)^{1/2}(\RR'_1)^{1/2}+\ep^2\De_0\c \OS'_{0,4}
 \eeaa
i.e.,
 \beaa
 \ep^{-1}\|\psi^{(s)}\|_{\Lsc^4(u,\ub)}&\les&\II_0'+\RR+\ep\De_0\c \OS'_{0,4}
\eeaa
On the other hand, for $s=1$, 
\beaa
 \|\chi\|_{\Lsc^4(u,\ub)}&\les& \|\chi\|_{\Lsc^4(u,0)}+ \| \a \|_{\Lsc(H_u)}^{1/2}  \|\nab\a   \|_{\Lsc(H_u)}^{1/2}        \\
 &+&\ep\De_0\c \|\chi\|_{\Lsc^4(u,\ub)}    +\ep^2\De_0\c \OS'_{0,4}      \\
 &\les& \|\chi\|_{\Lsc^4(u,0)}+\ep^{1/2}\RR+\ep\De_0\c \|\chi\|_{\Lsc^4(u,\ub)}    +\ep^2\De_0\c \OS'_{0,4}
\eeaa
i.e., for small enough $\ep$,
\beaa
 \ep^{-1/2}\|\chi\|_{\Lsc^4(u,\ub)}&\les&\ep^{-1/2} \|\chi\|_{\Lsc^4(u,0)}+\RR+
 \ep^{3/2}\De_0\c\OS'_{0,4}
\eeaa
\begin{proposition}
\label{prop:Ricciestim0.4}
Under    the bootstrap assumption $\OS_{0,\infty}\le \De_0$ and  assuming that $\ep \De_0$ is sufficiently small we derive,
\beaa
\OS_{0,4}[\omb,\eta]&\les& \II_0' + \RR +\ep \De_0\c \OS'_{0,4}\\
\OS_{0,4}[\trch]&\les&1+  \RR+  \ep \De_0 \c\OS_{0,4},\\
\OS_{0,4}[\chi] &\les&\RR+\ep\De_0\c\OS_{0,4}
\eeaa
\end{proposition}
 {\bf Remark}.   As in \cite{K-R:trapped} we can  get improved estimates for $\trch$, i.e. $\|\trch\|_{\Lsc^\infty}\les \ep^2$
 and  $\|\trch\|_{\Lsc^2}\les \ep$
 
\subsection{Estimates for  $\chib,\etab, \om$ }
The Ricci coefficients  $\etab, \chib$ and $\omb$ verify equations of the form
\bea
\nab_3 \psi^{(s)}=-\frac 1 2 k \, \trchb_0 \psi^{(s)}+\sum_{s_1+s_2=s}\psi^{(s_1)}\c\psi^{(s_2)}+\Psi^{(s)}
\label{eq:transp.symb3}
\eea
with $k$ a positive  integer.  If $s\ge 1/2$
 we have, after  a simple Gronwall inequality, (since $\Psi^{(s)}\neq \a$),
\beaa
\|\psi^{(s)}\|_{\Lsc^4(u,\ub)}&\les& \|\psi^{(s)}\|_{\Lsc^4(0,\ub)}+\int_0^u  \|\Psi^{(s)}\|_{\Lsc^4(u',\ub)}+  \ep^2 \De \c\OS'_{0,4}\nn\\
&\les& \|\psi^{(s)}\|_{\Lsc^4(0,\ub)}+\ep(\RRb'_0)^{\frac 12}(\RRb'_1)^{\frac 12} +\ep^2\De_0\c\OS_{0,4}
\eeaa
Hence,
\bea
\ep^{-1}\|\psi^{(s)}\|_{\Lsc^4(u,\ub)}&\les&\ep^{-1} \|\psi^{(s)}\|_{\Lsc^4(0,\ub)}+\RR+\ep\De_0\c\OS_{0,4}
\eea
To estimate $\chibh$ we use the estimate,
 \beaa
\nab_3\chibh&=&-\aa    +  \trchb_0\, \chibh      -\wtrchb\,  \chibh       -2\omegab \chibh
\eeaa
Thus,
  after a standard application of the Gronwall inequality,
\beaa
\|\chibh\|_{\Lsc^4( S_u)}&\les&\|\chibh\|_{\Lsc^4( S_0)}+
\int_0^u \|\aa\|_{\Lsc^4( S_{u'})}+\ep^2\c\De\c\OS'_{0,4}
\eeaa
i.e., 
\beaa
\|\chibh\|_{\Lsc^4( S_u)}&\les&\|\chibh\|_{\Lsc^4( S_0)}+\ep\RRb +\ep^2\c\De\c\OS'_{0,4}
\eeaa
Now observe that,
\beaa
\|\chibh\|_{\Lsc^4( S_0)}&\les&\ep^{1/2}\II^{(0)}
\eeaa
Indeed,  along $H_0$ (where $\om=0$), 
\beaa
\nab_4\chibh +\frac 1 2 \trch \chibh&=\nab\widehat{\otimes} \etab-\frac 12 \trchb \chih +\etab\widehat{\otimes} \etab
\eeaa
or,
\beaa
\nab_4\chibh &=&-\frac 1 2 \trchb_0 \chih+\nab\widehat{\otimes} \etab+\psi_g\c\psi
\eeaa
Hence,
\beaa
\|\chibh\|_{\Lsc^4(0,\ub)}&\les&\| \chih\|_{\Lsc^4(0,\ub)}+ \ep^{3/2} C
\eeaa
i.e.,
\beaa
\ep^{-1/2}\|\chibh\|_{\Lsc^4(0,\ub)}&\les&\ep^{-1/2}\| \chih\|_{\Lsc^4(0,\ub)}+ \ep^{} C\\
&\les& \II^{(0)}+ \ep^{} C
\eeaa
\section{Proof of theorem \ref{thm:mainII}}
We denote by   $\RRold$ amd $\OOold$ the curvature and Ricci coefficient norms   which can be obtained by formally choosing $\ep=\de^{1/2}$
in the definitions \eqref{RR-norms.f} and \eqref{OO-norms.f}.
These correspond precisely to the $\RR, \OO$ norms used
in  our paper \cite{K-R:trapped}.    Since we have assumed that the initial data quantity  $\RR^{(0)}$, defined in \eqref{RR-initial}, is uniformly distributed 
on the scale $\de^{1/2}\varpi^{-1}$,
\bea
\RRLasup^{(0)}\les \de^{\frac 12}\varpi^{-1} \RR^{(0)}
\eea
from which we also deduce, according to theorem \ref{thm.main.loc},
 \bea
 (\RRLasup+\RRbLasup)(u,\ub)\les \de^{\frac 12}\varpi^{-1} \RR^{(0)}
 \eea
Observe  that,  with  respect to the old scaling $\scold$  in  \cite{K-R:trapped},
we deduce, for all $0\le u\le 1$,
\beaa
\ep\| \fLa\a\|_{\Lscold(H_u)}+\|\fLa(\b, \ro, \si, \bb)\|_{\Lscold(H_u)}&\les&\ep\varpi^{-1} \RR^{(0)}\\
\|\fLa \nab \a\|_{\Lscold(H_u)}+\|\fLa \nab (\b, \ro, \si, \bb)\|_{\Lscold(H_u)}&\les&\ep\varpi^{-1} \RR^{(0)}
\eeaa
or, for  $ \ep\varpi^{-1}\les 1$
\begin{equation}
\begin{split}
\varpi  \|\fLa\a\|_{\Lscold(H_u)}+\|\fLa(\b, \ro, \si, \bb)\|_{\Lscold(H_u)}&\les \RR^{(0)}\\
\|\fLa \nab \a\|_{\Lscold(H_u)}+\|\fLa\nab (\b, \ro, \si, \bb\|_{\Lscold(H_u)}&\les \ep\varpi^{-1} \RR^{(0)}
\end{split}
\end{equation}

In particular,  if $\de^{1/2}\les \varpi$ 
 we deduce, relative to the old scaling $\scold$,
\bea
^{[\La]}\RRold(u,\ub)\les    \RR^{(0)}.\label{initial.curv.old}
\eea
with $^{[\La]}\RRold$ the localized version  of the norms 
$\RRold$, i.e. 
$
^{[\La]}\RRold=\sup_{\La}\, ^{(\La)}\RRold.
$

Moreover,  with   $ \ep\varpi^{-1}:=q$  a small parameter,
 we have, 
\bea
\sup_{\La}\|\fLa\nab (\b, \ro, \si, \bb)\|_{\Lscold(H_u)}&\les&q \label{initial.old.additional}
\eea
which, restricted to $u=0$,  is precisely  the localized version of  estimate (32) of proposition 2.8 in
\cite{K-R:trapped}.   In view of    theorem 2.6  in \cite{K-R:trapped},  the global version of  estimate \eqref{initial.curv.old},  i.e. $\RRold<\infty$,  allows one to deduce the boundedness of the global $\OOold$ norms, i.e.
 for some universal constant $C$,
$
\OOold\les C.
$
The global version of  condition \eqref{initial.old.additional}, with $q$ sufficiently small
 (which corresponds\footnote{With the small quantity $\ep$ replaced by $q$ here.} to condition (32) of proposition 2.8),  
  was necessary in the proof of theorem 2.7 in \cite{K-R:trapped} to insure the formation of a trapped surface.   The proof of formation of a trapped surface
 was based, in addition,  on  the  crucial  lower bound condition, 
$$
\int_0^\de |\chih_0|^2(\ub,\th) d\ub >\frac {2(r_0-u)}{r_0^2}.
$$
We note that  the proof of  formation of a trapped surface argument\footnote{See  the original argument in \cite{Chr:book}  and its  outline in the introduction
 to \cite{K-R:trapped}. },    is purely local in $\th$.  More precisely,  to show 
that $\trch(u,\de,\th)<0$  requires only the  control of  the Ricci coefficients in the 
domain\footnote{In actuality, because of the difference between the $\th$ and $\underline\theta$ coordinates 
defined respectively by parallel transport along $H_u$ and $\Hb_{\ub}$, the domain has to be enlarged 
to include all angles $\theta'$ such  that $|\theta'-\theta|\le \de^{\frac 12}$.} $\{(u',\ub,\th):\,
0\le u'\le u,\,\, 0\le\ub\le\de\}$. 
We further note that the global versions (unlocalized)  of  \eqref{initial.curv.old}, \eqref{initial.old.additional} make no reference to the support of 
the quantities involved and in particular are entirely compatible  with the possibility   that most or even all of the norm is 
concentrated in the support of  $\fLa$ for some specific $\La$.

In view of the above discussion we conclude that we could adapt  the 
proof  used in \cite{K-R:trapped}  for the formation of a trapped surface
to our situation provided that we  could   derive bounds for the localized
$\OOold$ norms from the boundedness of the  localized $\RRold$ norms. 
More
precisely we need to prove the following:
\begin{proposition}
Let $\{\La\}$ be a partition of $S_0$ of size $|\La|\approx \de^{\frac 12} q^{-1}$ with $q$ sufficiently small. Then
assuming that $^{[\La]}\OOold^{(0)}<\infty$,
$$
 ^{[\La]}\OOold\le C(^{[\La]}\OOold^{(0)},\,^{[\La]}\RRold, \, ^{[\La]}\RRbold)
$$
\end{proposition}

The proof of this proposition  is based on the observation
that all arguments used in sections  5-12 of \cite{K-R:trapped}
can be appropriately localized. This is particularly obvious
for those estimates which are derived  from transport equations.
Consider, for example, the transport  equations of the form
 \eqref{eq:transp.symb4} or,
simply, $
\nab_4\psi=\psi\c\psi +\Psi
$. Since $\nab_4 ^{(\La)} f=0$  and $\sum_\La \fLa=1$ we 
deduce,
\beaa
\nab_4\fLa \psi&=&\fLa \psi\c\psi +\fLa \Psi\\
&=&\sum_{\Lat} \fLa \psi\c \fLat   \psi +\fLa \Psi\
\eeaa
Hence, with respect to the old scaling,
\beaa
\|\fLa \psi\|_{\Lscold^4(u,\ub)}   &\les&\|\fLa \psi\|_{\Lscold^4(u,0)}+
\de^{-1}  \int_0^{\ub} \de^{1/2}
\sup_{\La} \|\fLa \psi\|_{\Lscold^\infty(u,\ub')}\sup_{\La}  \|\fLa \psi\|_{\Lscold^4(u,\ub') }d\ub'\\
&+& \de^{-1}  \int_0^{\ub} \de^{1/2}\sup_{\La} \|\fLa\Psi\|_{\Lscold^4(u,\ub')} d\ub'
\eeaa
Proceeding as in  \cite{K-R:trapped}  we  derive,
\beaa
\sup_{\La}\|\fLa \psi\|_{\Lscold^4(u,\ub)}  &\les&^{[\La]}  \RRold+\de^{1/2}\, ^{[\La]}\OOold_{0,\infty}   \c\, ^{[\La]}\OOold_{0,4} 
\eeaa

In the case of the transport equations of the form \eqref{eq:transp.symb3},
i.e., $\nab_3\psi=\psi\c\psi +\Psi$ we obtain,
$$
\nab_3 (\,^{(\La)}f\, \psi)= \,^{(\La)}f \psi\c\psi +\,^{(\La)}f \Psi + \ep \de^{\frac 12} |\La|^{-1}\, ^{(\La)}\tilde f\, \psi,
$$
where $\tilde f$ is a function similar to $f$ but with slightly large support. Using that $\de^{\frac 12} |\La|^{-1}\les q$
we easily obtain the estimate, for the corresponding Ricci components,
\beaa
\sup_{\La}\|\fLa \psi\|_{\Lscold^4(u,\ub)}  &\les&^{[\La]}  \RRbold
+\de^{1/2}\, ^{[\La]}\OOold_{0,\infty}   \c\, ^{[\La]}\OOold_{0,4}. 
\eeaa
The angular localization also affects the elliptic estimates for the Ricci coefficients. For the Codazzi equation
$$
\DD \psi=\psi\c\psi+\Psi
$$
we obtain 
\begin{align*}
\|\fLa\, \DD\psi\|_{\Lscold^2(u,\ub)}\les \|\fLa \psi\c\psi\|_{\Lscold^2(u,\ub)}+\|\fLa \Psi\|_{\Lscold^2(u,\ub)}\les
\de^{\frac 12}\, ^{[\La]}\OOold_{0,\infty}\c \, ^{[\La]}\OOold_{0,2} +  ^{[\La]}  \RRold_0 + ^{[\La]}  \RRbold_0
\end{align*}
On the other hand, integrating by parts and using the identity $\Delta=\DD^*\DD \pm K$ we obtain
\begin{align*}
\|\fLa\, \nab\psi\|_{\Lscold^2(u,\ub)}&\les \|\fLa\, \DD\psi\|_{\Lscold^2(u,\ub)}+  \|\nab \fLa\,\psi\|_{\Lscold^2(u,\ub)}
+ \|\fLa\, K\c\psi\|_{\Lscold^2(u,\ub)}\\ &\les  \|\fLa\, \DD\psi\|_{\Lscold^2(u,\ub)}+  \de^{\frac 12} |\La|^{-1}  \, ^{[\La]}\OOold_{0,2} + \de^{\frac 12}\, ^{[\La]} \RRold \, ^{[\La]}\OOold_{0,\infty} 
\end{align*}

 \end{document}